\documentclass[12pt,a4paper,final]{iopart}
  
\usepackage{iopams}  
\usepackage{graphicx}

\usepackage[breaklinks=true,colorlinks=true,linkcolor=blue,urlcolor=blue,citecolor=blue]{hyperref}
\usepackage[export]{adjustbox}
\usepackage{hyperref}
\usepackage{graphicx}
\usepackage{color}

\newcommand{\mbf}[1]{\mathbf{#1}}
\newcommand{\heta}{\boldsymbol{\eta}}

\newcommand{\ba}{b^{\phantom{\dagger}}}
\newcommand{\bc}{b^\dagger}

\newcommand{\LW}{\hat{\Phi}^{\rm{\tiny (LW)}}_V}

\newcommand{\BK}{\hat{\Gamma}^{\rm{\tiny (BK)}}_{\bF \mbf{t} \heta V}}

\newcommand{\SE}{\hat{\Gamma}^{\rm{\tiny (SE)}}_{\bF \mbf{t} \heta V}}
\newcommand{\ASE}{\hat{\Gamma}^{\rm{\tiny (SE)}}_{\bF \mbf{t} P V}}
\newcommand{\ASEprime}{\hat{\Gamma}^{\rm{\tiny (SE)}}_{\bF' \bDelta P V}}
\newcommand{\SFT}{\hat{\Gamma}^{\rm{\tiny (SFT)}}_{\bF \mbf{t} P V}}

\newcommand{\ALWSE}{\hat{\mathcal{T}}_{P V}}
\newcommand{\LWSE}{\hat{\mathcal{F}}_V}

\newcommand{\n}{\hat{n}}

\renewcommand{\Tr}{\rm{Tr}}

\newcommand{\astcycl}{\mathrlap{\kern0.085em{\circlearrowright}}\ast}

\renewcommand{\S}{\mathcal{S}}

\newcommand{\taucycl}{\mathrlap{\kern0.42em{\bullet}}\circlearrowright}

\newcommand{\bbb}{\mbf{b}}
\newcommand{\bba}{\mbf{b}^{\phantom{\dagger}}}
\newcommand{\bbc}{\mbf{b}^{\dagger}}

\newcommand{\bG}{\mbf{G}}

\newcommand{\bF}{\mbf{F}}
\newcommand{\bt}{\mbf{t}}
\newcommand{\bS}{\boldsymbol{\Sigma}}

\renewcommand{\bPhi}{\boldsymbol\Phi}
\newcommand{\bP}{\boldsymbol\Phi}

\renewcommand{\bDelta}{\boldsymbol{\Delta}}

\newcommand{\bGeta}{\mbf{G}_{\heta}}

\newcommand{\bPeta}{\boldsymbol{\Phi}_{\heta}}

\newcommand{\ranp}{\rangle_{P}}

\begin{document}
% ----------------------------------------------------------------------

\title{Self-energy functional theory with symmetry breaking for disordered lattice bosons}

%\title{Competition between disorder-driven delocalization and localization of lattice bosons in a self-energy functional approach}

\author{Dario H\"ugel}
%\email{}
\address{Department of Physics, Arnold Sommerfeld Center for Theoretical Physics and Center for NanoScience, Ludwig-Maximilians-Universit\"at M\"unchen, Theresienstrasse 37, 80333 Munich, Germany} 

\author{Hugo U.~R.~Strand}
\address{Department of Quantum Matter Physics, University of Geneva, 24 Quai Ernest-Ansermet, 1211 Geneva 4, Switzerland} 
\address{Center for Computational Quantum Physics, Flatiron Institute, 162 Fifth Avenue, New York, NY 10010, USA} 

\author{Lode Pollet}
%\email{philipp.werner@unifr.ch}
\address{Department of Physics, Arnold Sommerfeld Center for Theoretical Physics and Center for NanoScience, Ludwig-Maximilians-Universit\"at M\"unchen, Theresienstrasse 37, 80333 Munich, Germany} 

\date{\today} 
%\pacs{71.10.Fd, 05.30.Jp, 05.30.-d, 05.50.+q}

% Hubbard model electronic structure, 71.10.Fd
% Boson systems, 05.30.Jp
% Bose-Einstein statistics, 05.30.-d
% Lattice theory and statistics, 05.50.+q

% Boson systems, 05.30.Jp
% Bose-Einstein statistics, 05.30.-d
% Ultracold gases, 67.85.-d

%03.75.Hh Static properties of condensates; thermodynamical, statistical, and structural properties
%03.75.Kk Dynamic properties of condensates; collective and hydrodynamic excitations, superfluid flow
%05.30.Jp Boson systems 
%05.30.Rt Quantum phase transitions
%05.70.Ln Nonequilibrium and irreversible thermodynamics
%37.10.Jk Atoms in optical lattices
%64.60.Ht Dynamic critical phenomena
%67.10.-j Quantum fluids: general properties
%67.10.Ba Boson degeneracy
%67.85.De Dynamic properties of condensates; excitations, and superfluid flow
%67.85.Hj Bose-Einstein condensates in optical potentials
%71.10.Fd Lattice fermion models (Hubbard model, etc.)
%71.27.+a Strongly correlated electron systems; heavy fermions

% --------------------------------------------------------------------
\begin{abstract}
  We extend the self-energy functional theory (SFT) to the case of interacting lattice bosons in the presence of symmetry breaking and quenched disorder.
  The self-energy functional we derive depends only on the self-energies of the disorder-averaged propagators, allowing for the construction of general non-perturbative approximations.
  Using a simple single-site reference system with only three variational parameters, we are able to reproduce numerically exact quantum Monte Carlo (QMC) results on local observables of the 
  Bose-Hubbard model with box disorder with high accuracy. At strong interactions, the phase boundaries are reproduced qualitatively but shifted with respect to the ones observed with QMC due to the extremely low condensate fraction in the superfluid phase.
  Deep in the strongly-disordered weakly-interacting regime, the simple reference system employed is insufficient and no stationary solutions can be found within its restricted variational subspace. 
  By systematically analyzing thermodynamical observables and the spectral function, we find that the strongly-interacting Bose glass is characterized by different regimes, depending on which local occupations are activated as a function of the disorder strength.
 We find that the particles delocalize into isolated superfluid lakes over a strongly localized background around maximally-occupied sites whenever these sites are particularly rare.
  Our results indicate that the transition from the Bose glass to the superfluid phase around unit filling at strong interactions is driven by the percolation of superfluid lakes which form around doubly occupied sites.
\end{abstract}
% --------------------------------------------------------------------

\maketitle
\makeatletter
\let\toc@pre\relax
\let\toc@post\relax
\makeatother

% ----------------------------------------------------------------------
\section{Introduction}
% ----------------------------------------------------------------------

% -- General introduction to the field
%
Ever since the seminal work by Giamarchi and Schulz~\cite{GiamarchiSchulz1987,GiamarchiSchulz1988} in one dimension and the extension to any dimension by Fisher \textit{et al.~}\cite{MF_Fisher}, the intricate interplay of disorder and interactions in bosonic lattice systems has been an active field of research.
The advent of cold atom experiments in optical lattices \cite{Morsch:2006vn, BlochRev}, where disorder can be realized e.g.\ through the overlap of optical potentials with incommensurate wavelengths \cite{Fellani_07,Roati_08} or speckle-laser patterns \cite{Chen_Speck,Speck_Lie,Speck_Pasi},
has further invigorated the interest in this class of systems.
More recently, the field has moved to research frontiers such as many-body-localization \cite{Huse_MBL} thanks to advances in monitoring real-time dynamics and state preparation.

% -- Numerical methods, exact methods limited to finite systems
%
The theoretical understanding of disordered and interacting lattice bosons has been primarily driven by numerical simulations.
Exact diagonalization (ED) \cite{Runge_92} can only be applied to relatively small finite system sizes, while the extension of the density matrix renormalization group (DMRG) \cite{Kuhner, Rapsch, Kollath, Kollath2} to disorder \cite{Rapsch_99} is restricted to low-dimensional systems.
In higher dimensions the state of the art method is
path integral quantum Monte Carlo (QMC) with worm updates \cite{Tupitsyn,Trotzky,Rev_Lode,Gur_09}. This algorithm provides numerically exact results for large but finite-sized bosonic lattice systems, while the disorder can be accounted for by averaging over many disorder realizations \cite{Gur_09}.
However, dynamical quantities such as the single-particle spectral function can only be determined by performing analytic continuation of imaginary-time propagators with stochastic noise \cite{Jarrell:1996fj, Pippan:2009aa}. The continuation is an inherently ill-posed problem, and cannot resolve sharp resonances.
While the methods mentioned above excel with a high numerical accuracy, they rely on finite system sizes, which can represent a problem when rare disorder-driven fluctuations play an important role, which can only be captured once one approaches the thermodynamical limit.
%
% -- Mean field methods
%
The available methods in the thermodynamical limit rely on approximations.
The mean-field decoupling approximation \cite{MF_Fisher} can be applied to disordered systems using an arithmetically averaged condensate. However, this overestimates the phase coherence and the extent of the superfluid phases, as locally condensed bosons are mistaken for a global condensate \cite{Arithm_Dis}. In fact, in mean-field methods with position-space resolution the geometric percolation of condensed regions appears to be a more accurate quantity to evaluate the global superfluid response \cite{Rieger_13}.
A more accurate mean-field approach is the stochastic mean-field theory \cite{Stoch_MF0,Stoch_MF}, where the condensate order parameter is treated as a disorder-dependent quantity. However, mean-field methods are self-consistent only in terms of the condensate (i.e.\ the one-point propagator). This severely hampers the ability to describe uncondensed phases, which are simply approximated by the zero hopping limit (i.e.\ the atomic limit).
%%

% -- Dynamical mean-field theory
%
A non-perturbative method which includes also a self-consistency in terms of the two-point propagator is the dynamical mean-field theory (DMFT), originally formulated for fermions \cite{Metzner:1989aa, Georges:1996aa} and later generalized to bosons \cite{Byc_BDMFT, Hu:2009qf, Hubener:2009cr, BDMFT, BDMFT1, Panas:2015ab}.
For fermions the formalism has been extended to disordered systems  \cite{Dobro_PRL_97,Dobro_PRL_03,Byc_PRL_05,Byc_PRB_05} by averaging the systems propagators over all disorder configurations.
While an arithmetic averaging in this framework works well for weak disorder, it misses the essential physics in non-self-averaging phases. In such phases, like the Anderson-localized regime \cite{Anderson_58,Dobro_PRL_03}, observables show broad tails in their disorder-distribution.
An interesting idea for incorporating non-self-averaging effects is the typical medium theory \cite{Dobro_PRL_97,Dobro_PRL_03,Byc_PRL_05,Byc_PRB_05}, where the arithmetic average is replaced by a geometrical mean. However, it is not clear what the range of validity is for this approach.
We would like to point out that the works above employing disorder and DMFT all study fermionic systems. As of today we are not aware of any works applying DMFT to disordered \emph{bosonic} systems.

A more general theoretical framework for constructing non-perturbative approximations for interacting many-body systems is the self-energy functional theory (SFT) 
\cite{Potthoff:2003aa, Potthoff:2003ab, Potthoff:2006aa, Springer:2012,Koller:2006aa,Knap:2010aa,Arrigoni:2011aa,BSFT}, from which DMFT can be derived as a certain constriction of the variational space.
The formalism was first developed for fermions \cite{Potthoff:2003aa, Potthoff:2003ab, Potthoff:2006aa, Springer:2012} and later extended to bosonic systems \cite{Koller:2006aa, Knap:2010aa, Arrigoni:2011aa, BSFT}. Our recent derivation \cite{BSFT} based on the bosonic Baym-Kadanoff functional \cite{De-Dominicis:1964aa, De-Dominicis:1964ab} correctly includes $U(1)$-symmetry-breaking, and simplifies to bosonic DMFT \cite{Byc_BDMFT, Hu:2009qf, Hubener:2009cr, BDMFT, BDMFT1, Panas:2015ab} in a particular limit.
Within SFT,
non-perturbative approximations are readily constructed by
restricting the self-energy domain of the original lattice system to the self-energies of a simpler exactly solvable \textit{reference} system.
This reduces the full complexity of the original problem to a search for stationary solutions in terms of the variational free propagators of the reference system.
The generalization of SFT to systems with disorder has been developed for fermions in Ref.\ \cite{Pot_Dis} and applied in a variational cluster approximation to bosons in the absence of $U(1)$-symmetry-breaking in Ref.\ \cite{Kanp_Disorder}.

% -- The purpose of this paper
%
The aim of this paper is to extend the bosonic SFT formalism of Ref.\ \cite{BSFT} to disordered lattice bosons including the possibility of $U(1)$-symmetry-breaking. As argued for fermions in Ref.\ \cite{Pot_Dis}, the geometrical mean used in the context of DMFT \cite{Dobro_PRL_97,Dobro_PRL_03,Byc_PRL_05,Byc_PRB_05} is hard to reconcile with the variational SFT framework. We therefore derive an arithmetically averaged formalism, where, through the introduction of an appropriate $\ALWSE$ functional, the functional depends only on the self-energies of the arithmetically averaged propagators. Just as the version for clean systems \cite{BSFT}, we find that SFT incorporates a disorder-averaged generalization of bosonic DMFT \cite{Byc_BDMFT, Hu:2009qf, Hubener:2009cr, BDMFT, BDMFT1, Panas:2015ab} in a certain limit. The resulting functional is, however, more general than DMFT by being amenable to a more general variational space.

% -- Bose-Hubbard model with disorder, Bose Glass
%
The prototypical model for interacting disordered bosonic lattice systems is the Bose-Hubbard model (BHm) in the presence of local disorder (for a review, see Ref.\ \cite{Pollet_CRP}). In addition to the Mott insulating and superfluid phases of the clean system, the groundstate phase diagram exhibits a new phase: the Bose glass \cite{GiamarchiSchulz1987, GiamarchiSchulz1988,MF_Fisher}. This is an insulating but gapless and compressible phase, which always intervenes between the Mott insulator and the superfluid phase at finite disorder  \cite{Svistunov1996,Incl_Dis,Gur_09}.  While certain single-particle states can show a high (but not macroscopic) occupation, the disorder does not allow for global phase coherence as observed in the superfluid. The statistical fluctuations of the local potential, on the other hand, locally exceed the gaps of adding/removing a particle \cite{Gur_09}, creating gapless regions which induce a non-vanishing density of states at zero energy \cite{MF_Fisher}. 
%
% -- dBHm ground state phase diagram
%
In the groundstate phase diagram of the disordered BHm on a cubic lattice the superfluid phase extends to surprisingly large values of the interaction $U$ and the disorder strength $D$. For low/intermediate interactions and high disorder it can be argued that this is related to the percolation between localized states \cite{Gur_09}. At stronger interaction and lower disorder, the phase diagram is characterized by the so-called ``superfluid finger'' which extends to much larger interactions than  the critical value of the clean system and is characterized by an extremely low condensate fraction. The critical temperature at which the condensate vanishes is thus extremely low, making it very hard to access this regime in experiments \cite{Gur_09}.

% -- Summary of salient results
%
We apply SFT to the BHm with local box disorder on the cubic lattice using the simplest imaginable reference system, comprising a single bosonic mode. This restriction to the minimal reference system produces a self-energy approximation with three variational degrees of freedom, which we will denote by SFA3. In the clean BHm the SFA3 approach has been shown to be in quantitative agreement with numerically exact QMC results \cite{BSFT}.
%
%the so-called SFA3 reference system, consisting of just three variational degrees of freedom yielding excellent agreement with QMC in the clean BHm \cite{BSFT}.
%
In this work we investigate the disordered BHm in the vicinity of the superfluid finger, where the condensate density is extremely fragile, leading to a substantial shift in the phase boundaries even if the numerical error is very low. Nonetheless, we observe excellent agreement of the thermodynamic quantities computed with SFT and the QMC reference results.

% -- New physical understanding of the Bose glass
%
Since the SFA3 reference system can be solved exactly, we can also evaluate the lattice spectral function and thereby obtain spectroscopic information not readily available from QMC.
By systematically analyzing the local excitations of the SFA3 spectral functions, we find that the strongly-interacting Bose glass is characterized by different regimes, depending on which local occupations $n$ are activated as a function of the disorder strength $D$. While local observables are described well by the atomic limit, we find that the particles delocalize into isolated superfluid lakes over the strongly-localized background around highly-occupied sites whenever these sites are particularly rare. In particular, our results indicate that the transition from the strongly interacting Bose glass to the superfluid phase close to unit filling is driven by the percolation of superfluid lakes which form around doubly occupied sites. As $D$ is further increased and the number of highly-occupied sites increases accordingly, the particles are localized by the increasing particle-number fluctuations and interaction energy, explaining the reentrant behavior of the superfluid finger at larger $D$.

We also present results deeper in the superfluid phase (i.e.\ at weaker interactions), showing excellent agreement with QMC for thermodynamical quantities at low disorder.
%
%As the disorder further increases and
%
When the disorder
dominates both over the bandwidth and the interaction,
%
%our approach of using a
%
 the restricted variational subspace of our SFA3 reference system is however insufficient, as we no-longer can stabilize a stationary solution.
Whether this can be remedied by a more general reference system construction is an open question.

This paper is organized as follows. In Sec.\ \ref{sec:theory} we derive the self-energy functional theory for disordered lattice bosons: starting from the free-energy functional (Sec.\ \ref{subsec:FreeEnergyFunctional}), we generalize the bosonic Baym-Kadanoff functional to the case of disorder (Sec.\ \ref{sec:BK}), perform a Legendre transform to the self-energy effective action (Sec.\ \ref{sec:SE}), average the effective action over all disorder configurations (Sec.\ \ref{sec:AvgSE}), and finally arrive at the disorder-averaged self-energy functional (Sec.\ \ref{sec:Avg_SFT}). In Sec.\ \ref{sec:model} we introduce the disordered BHm, discuss the SFA3 reference system used in the SFT calculations (Sec.\ \ref{sec:SFA3}) and derive analytic results in the atomic limit (Sec.\ \ref{sec:atomic_limit}). The numerical results are presented in Sec.\ \ref{sec:results}, where we investigate the strongly-interacting Bose glass (Sec.\ \ref{sec:strong_BG}), the strongly-interacting superfluid phase transition (Sec.\ \ref{sec:SF_Finger}), and the superfluid phase (Sec.\ \ref{sec:SF_phase}). Finally, Sec.\ \ref{sec:conc} is devoted to the conclusion.

% ----------------------------------------------------------------------
\section{Self-energy functional theory for disordered lattice bosons}
\label{sec:theory}
% ----------------------------------------------------------------------

In this section we derive the self-energy functional theory for disordered lattice bosons. In analogy to the formalism for clean systems derived in Ref.\ \cite{BSFT}, we do so by a series of Legendre transformations starting from the free-energy functional and introduce a simpler exactly solvable reference system sharing the same local interaction and disorder distribution. As was done in a previous work on fermions \cite{Pot_Dis}, we average over all possible disorder configurations, arriving at a functional which only depends on the self-energies of the arithmetically averaged propagators of the system.

Note that, in order to keep track of the various additional dependencies arising through the disorder, we introduce a slightly more complex notation than in our work on disorder-free bosons in Ref.\ \cite{BSFT}, by denoting the explicit dependencies on system parameters as subscripts. Further, we will denote Nambu objects (i.e.\ matrices or vectors) as bold letters (e.g.\ $\mbf{O}$), scalars as simple letters  (e.g.\ $O$), and functionals with a hat (e.g.\ $\hat{O}$). Finally, for notational simplicity, we denote the one-point self-energy (formerly $ \mbf{\Sigma_{1/2}}$ in Ref.\ \cite{BSFT}) as $\mbf{S}$.

% ----------------------------------------------------------------------
\subsection{Free-energy functional}
\label{subsec:FreeEnergyFunctional}
% ----------------------------------------------------------------------

Consider a lattice system of bosons in the presence of quadratic disorder, with creation (annihilation) operator $\bc_i$ ($\ba_i$) on site $i$.
Using the Nambu operators $\bbc_\alpha \equiv \bbc_{i\nu} \equiv (\bc_i, \ba_i)_\nu$ with commutator $[\bbb^{\alpha}, \bbc_\beta] = (\mbf{1} \otimes \sigma_z)^\alpha_\beta$ , where $\alpha$ is a superindex spanning both the site index $i$ and the Nambu index $\nu$, the Hamiltonian $\hat{H}$ of the system can be written as
\begin{equation}
  \hat{H} =
  \bF_\alpha^\dagger \bbb^\alpha
  + \frac{1}{2}\bbc_\alpha \bt^{\alpha}_{\beta} \bbb^\beta
  +\frac{1}{2} \bbc_\alpha \heta^{\alpha}_{\beta} \bbb^\beta
  + \hat{V}
  %+ \hat{H}_{0}
  \, ,
  \label{eq:H}
\end{equation}
where repeated indices are summed over, $\bF$ is an explicit symmetry-breaking field, $\bt^{\alpha}_\beta = \bt^{i\eta}_{j\nu} = t_{ij} \otimes \mbf{1}_{\eta\nu}$ is the hopping, the quadratic disorder $\heta^\alpha_\beta = \heta^{i\eta}_{j\nu} = \eta_{ij} \otimes \mbf{1}_{\eta\nu}$ has the probability distribution $P(\heta)$, and the general three and four-body interaction $\hat{V}$ can be expressed as $\hat{V} \equiv V^{(3)}_{\alpha\beta\gamma} \bbb^\alpha \bbb^\beta \bbb^\gamma + V^{(4)}_{\alpha\beta\gamma\delta} \bbb^\alpha \bbb^\beta \bbb^\gamma \bbb^\delta$.
To keep the notation compact we will henceforth suppress the lattice-Nambu superindices. 

At finite temperature $T \equiv \beta^{-1}$ the free energy functional of the interacting system is given by
\begin{equation}
  \hat{\Omega}_{V}[\bF, \bG^{-1}_0] = - \ln(\Tr[e^{-\S_{V}[\bF, \bG^{-1}_0]}])/\beta
  \label{eq:Omega} \, ,
\end{equation}
where the subscript '$V$' means that in addition to $\bF$ and $\bG^{-1}_0$ the functional also depends on the interaction vertex $\hat{V}$, and $\S_{V}$ is the imaginary-time action
\begin{eqnarray}
  \S_{V}[\bF, \bG^{-1}_0] \equiv &
  \int_0^\beta d\tau \,
  \bF^\dagger \bbb(\tau)
  + \int_0^\beta d\tau \,
  \hat{V}[\bbb(\tau)] \nonumber
  \\
  &- \frac{1}{2} \int\int_0^\beta d\tau d\tau' \,
  \bbc(\tau)  \bG_0^{-1}(\tau, \tau')\bbb(\tau')
  \, .
\end{eqnarray}
The free energy functional $\hat{\Omega}_{V}[\bF, \bG^{-1}_0]$ is equal to the free energy $\Omega_{\bF \mbf{t} \heta V}$ of the lattice system in Eq.\ (\ref{eq:H}) with fix disorder configuration $\heta$, when evaluated at the symmetry breaking field $\bF$ and the free single-particle propagator $\bG_{\mbf{t} \heta 0}$, i.e.
\begin{equation}
  \hat{\Omega}_V[\bF, \bG_{\mbf{t} \heta 0}^{-1}] = \Omega_{\bF \mbf{t} \heta V}
  \, , \nonumber
\end{equation}
where the non-interacting ($\hat{V}=0$) single-particle propagator of Eq.\ (\ref{eq:H}) is given by
\begin{equation}
  \bG_{\mbf{t} \heta 0}^{-1}(\tau, \tau')
  =
    \delta(\tau - \tau')
  (- [\mbf{1} \otimes \sigma_z] \partial_{\tau'} - \bt - \heta)
    \, ,
    \label{eq:SingleParticlePropagator}
\end{equation}
and the subscript means that it depends on the hopping $\mbf{t}$ and the disorder configuration $\heta$ only.
By taking functional derivatives of the free energy functional $\hat{\Omega}_V$ with respect to $\bF$ and $\bG_0^{-1}$ we obtain the two functionals
\begin{eqnarray}
  \hat{\phi}_V[\bF, \bG_0^{-1}]
  & \equiv
  \beta \frac{\delta \hat{\Omega}_V[\bF, \bG_0^{-1}]}{\delta \bF^\dagger},
  \label{eq:Ffunc}
  \\
  \hat{\mathcal{G}}_V[\bF, \bG_0^{-1}]
  & \equiv
  2\beta \frac{\delta \hat{\Omega}_V[\bF, \bG_0^{-1}]}{ \delta \bG_0^{-1}}
  +
  \big(
  \hat{\phi}_V
  \hat{\phi}_V^\dagger
  \big)[\bF, \bG_0^{-1}],
  \label{eq:Gfunc}
\end{eqnarray}
that reproduce the physical one- and two-point propagators (i.e. the condensate $\bPhi_{\bF \mbf{t} \heta V}$ and the connected Green's function $\bG_{\bF \mbf{t} \heta V}$) of the disordered interacting system in Eq.\ (\ref{eq:H}) when evaluated at $\bF$ and $\bG_{\mbf{t} \heta 0}^{-1}$, i.e.
\begin{eqnarray}
  \hat{\phi}_V[\bF, \bG_{\mbf{t} \heta 0}^{-1}] & = \bPhi_{\bF \mbf{t} \heta V}=\langle \mbf{b}\rangle
  \, , \nonumber\\
  \hat{\mathcal{G}}_V[\bF, \bG_{\mbf{t} \heta 0}^{-1}] & = \bG_{\bF \mbf{t} \heta V}=-\langle \mbf{b}(\tau)\bbc(0)\rangle+\langle \mbf{b}\rangle\langle \bbc\rangle
  \, .
\end{eqnarray}
%

% ----------------------------------------------------------------------
\subsection{Baym-Kadanoff functional}
\label{sec:BK}
% ----------------------------------------------------------------------

When exchanging the functional dependence of the free energy functional $\hat{\Omega}_V$ in Eq.\ (\ref{eq:Omega}), from $\bF$ and $\bG_0^{-1}$ to $\bPhi$ and $\bG$ by means of a Legendre transformation, one obtains the bosonic Baym-Kadanoff functional  \cite{De-Dominicis:1964aa,De-Dominicis:1964ab,BSFT}
\begin{eqnarray} 
  \BK[\bPhi, \bG]
  =&
  \bF^\dagger \bPhi - \frac{1}{2} \bPhi^\dagger \bG^{-1}_{\mbf{t} \heta 0} \bPhi
  + \frac{1}{2} \Tr[ \bG^{-1}_{\mbf{t} \heta 0} \bG ]   \nonumber\\
  &+ \frac{1}{2} \Tr \ln [ -\bG^{-1} ]
  + \LW[\bPhi, \bG]\, .
\end{eqnarray}
Here, the entire complexity of the many-body system is contained in the Luttinger-Ward functional $ \LW[\bPhi, \bG]$ which contains all two-particle irreducible diagrams \cite{Luttinger:1960aa,Kleinert:1982aa}. For a more detailed discussion of the Luttinger-Ward functional in the context of SFT, see Ref.\ \cite{BSFT}.

At the physical interacting one and two-point propagators, the Baym-Kadanoff functional $\BK$ is stationary
\begin{equation}
  \partial_{\bPhi} \BK
  [\bPhi_{\bF \mbf{t} \heta V}, \bG_{\bF \mbf{t} \heta V}]  = 0 \, ,
\quad
  \partial_{\bG} \BK
  [\bPhi_{\bF \mbf{t} \heta V}, \bG_{\bF \mbf{t} \heta V}]  = 0 \, ,
  \label{eq:BKstat}
\end{equation}
and equal to the free energy
\begin{equation}
  \BK
  [\bPhi_{\bF \mbf{t} \heta V}, \bG_{\bF \mbf{t} \heta V}]
  = \Omega_{\bF \mbf{t} \heta V}
  \, .\nonumber
\end{equation}
The explicit functional derivatives take the form
\begin{equation}
  \frac{\delta \BK}{\delta \bPhi^\dagger}
   =
  \bF - \bG^{-1}_0 \bPhi
  + \frac{\delta \LW}{\delta \bPhi^\dagger}
  %  - \mbf{S}
  \, , \quad
  2\frac{\delta \BK}{\delta \bG}
   =
  \bG^{-1}_0 - \bG^{-1}
  + 2 \frac{\delta \LW}{\delta \bG}
  %- \bS
  \, . \nonumber
\end{equation}
By identifying the variations of the Luttinger-Ward functional $\LW$ as the one and two-point self-energies \cite{De-Dominicis:1964aa,De-Dominicis:1964ab,BSFT}
\begin{equation}
  \mbf{S} = - \delta_{\bPhi^\dagger} \LW
  \, , \quad
  \bS = -2 \delta_\bG \LW
  \, , \nonumber
\end{equation}
and applying the stationarity conditions [Eq.\ (\ref{eq:BKstat})] we find that the interacting propagators fulfill the two Dyson equations
\begin{equation}
  \bG_0^{-1} \bPhi  = \bF - \mbf{S} \, ,
 \quad
  \bG^{-1}  = \bG_0^{-1} - \bS \, .
  \label{eq:D1}
\end{equation}

Consider now the result of substituting $\bF$ and $\bG_0^{-1}$ using Eq.\ (\ref{eq:D1}) in the functionals $\hat{\phi}_V$ and $\hat{\mathcal{G}}_V$ [Eqs.\ (\ref{eq:Ffunc}) and (\ref{eq:Gfunc})]. This gives the highly non-linear coupled equations
\begin{eqnarray}
  \hat{\phi}_V
  [ (\bG^{-1} + \bS)\bPhi + \mbf{S}, \bG^{-1} + \bS]
  = \bPhi
  \, , \label{eq:nonlin1} \\
  \hat{\mathcal{G}}_V
  [ (\bG^{-1} + \bS)\bPhi + \mbf{S}, \bG^{-1} + \bS]
  = \bG
  \, . \label{eq:nonlin2}
\end{eqnarray}
For given self-energies $\mbf{S}$ and $\bS$ the concomitant solution of Eqs. (\ref{eq:nonlin1}) and (\ref{eq:nonlin2}) implicitly defines the functionals
\begin{equation}
  \hat{\bPhi}_V[\mbf{S}, \bS] = \bPhi
  \, , \quad
  \hat{\bG}_V[\mbf{S}, \bS] = \bG,
  \nonumber
\end{equation}
depending solely on the self-energies $\mbf{S}$ and $\bS$ and the interaction $\hat{V}$, producing the physical interacting propagators when evaluated at the physical self-energies, i.e.
\begin{equation}
  \hat{\bPhi}_V[\mbf{S}_{\bF \mbf{t} \heta V}, \bS_{\bF \mbf{t} \heta V}]  = \bPhi_{\bF \mbf{t} \heta V}
  \, , \quad
  \hat{\bG}_V[\mbf{S}_{\bF \mbf{t} \heta V}, \bS_{\bF \mbf{t} \heta V}]  = \bG_{\bF \mbf{t} \heta V} \label{eq:G_funct_eta}
  \, .
\end{equation}

% ----------------------------------------------------------------------
\subsection{Bosonic self-energy effective action}
\label{sec:SE}
% ----------------------------------------------------------------------

By means of a further Legendre transform the Baym-Kadanoff functional $\BK$ with functional dependence on $\bPhi$ and $\bG$ can be transformed into the self-energy effective action
\begin{equation}
  \SE[\mbf{S}, \bS]
  =
  \frac{1}{2} (\bF - \mbf{S})^\dagger \bG_{ \mbf{t} \heta 0} (\bF - \mbf{S})
  + \frac{1}{2} \Tr \ln [ - (\bG^{-1}_{\mbf{t} \heta 0} - \bS)]
  + \LWSE[\mbf{S}, \bS],
  \label{eq:SE_full}
\end{equation}
depending on the self-energies $\mbf{S}$ and $\bS$, where the universal functional $\LWSE[\mbf{S}, \bS]$ is the Legendre transform of the universal Luttinger-Ward functional $\LW[\bPhi, \bG]$, with variations (see Ref.\ \cite{BSFT} for details)
\begin{equation}
  \delta_{\mbf{S}^{\dagger}} \LWSE = \bPhi
  \, , \quad
  \delta_{\bS} \LWSE = \bG. \label{eq:vars_F}
\end{equation}
The variations of the self-energy effective action give
\begin{equation}
  \frac{\delta \SE}{\delta \mbf{S}^{\dagger}}  = - \bG_{\mbf{t} \heta 0}(\bF - \mbf{S}) + \bPhi
  \, ,\quad
  \frac{\delta \SE}{\delta \bS}  = - \left[\bG_{\mbf{t} \heta 0}^{-1} - \bS\right]^{-1} + \bG
  \, ,
\end{equation}
whence $\SE$ is stationary at the physical self-energies
\begin{equation}
  \delta_{\mbf{S}^{\dagger}} \SE[\mbf{S}_{\bF \mbf{t} \heta V}, \bS_{\bF \mbf{t} \heta V}]  = 0
  \, , \quad
  \delta_{\bS} \SE[\mbf{S}_{\bF \mbf{t} \heta V}, \bS_{\bF \mbf{t} \heta V}]  = 0
  \, .
\end{equation}
and equal to the free energy
\begin{equation}
  \SE[\mbf{S}_{\bF \mbf{t} \heta V}, \bS_{\bF \mbf{t} \heta V}] = \Omega_{\bF \mbf{t} \heta V} \nonumber
  \, .
  \label{eq:SE_physical}
\end{equation}

% ----------------------------------------------------------------------
\subsection{Disorder-averaged self-energy effective action}
\label{sec:AvgSE}
% ----------------------------------------------------------------------

While we up till now have treated a system with a single disorder realization $\heta$, we are interested in describing the averaged ensemble of systems with disorder probability distribution $P(\heta)$ and its ensemble averaged free-energy
\begin{equation} 
  \Omega_{\bF \mbf{t} P V}
  \equiv \langle \Omega_{\bF \mbf{t} \heta V} \rangle_P \equiv \int d\heta  P(\heta) \Omega_{\bF \mbf{t} \heta V}
  \, .
  \label{eq:EnsembleAverage}
\end{equation}
In terms of the self-energy functional Eq.\ (\ref{eq:SE_full}) the averaged free energy can be expressed as
\begin{equation}
  \Omega_{\bF \mbf{t} P V}
  =
  \left\langle
    \SE[\mbf{S}_{\bF \mbf{t} \heta V}, \bS_{\bF \mbf{t} \heta V}]
  \right\rangle_P
  \, . \nonumber
  \label{eq:DisFreeEnergy_in_SE}
\end{equation}
using Eq.\ (\ref{eq:SE_physical}). However, a direct application of the avaraged self-energy functional does not lend itself to the construction of approximations using disorder averaged propagators and self-energies.

% --------------------------------------------------------------------
%\textcolor{blue}{
% --------------------------------------------------------------------
%To this end we define the averaged self-energy effective action
%
%\begin{equation}
%  \ASE[\{ \mbf{S}_{\heta}, \bS_{\heta} \} ]
%  \equiv
%  \left\langle \SE[\mbf{S}_{\heta}, \bS_{\heta}] \right\rangle_P
%  \, . \nonumber
%\end{equation}
%
%where $\mbf{S}_{\heta}$ and $\bS_{\heta}$ denote the self-energies for the disorder configuration $\heta$.
% --------------------------------------------------------------------
%} % blue
% --------------------------------------------------------------------

To describe the combined effect of disorder and interaction we seek to construct an extended disorder-averaged functional parametrized by the disorder-averaged propagators
\begin{equation}
  \bar{\bPhi}
  \equiv
  \left\langle \hat{\bPhi}_{\heta} \right\rangle_P
  \, , \quad
  \bar{\bG}
  \equiv
  \left\langle
    \hat{\bG}_{\heta}
    - \hat{\bPhi}_{\heta} \hat{\bPhi}_{\heta}^\dagger
  \right\rangle_P
  + \bar{\bPhi} \bar{\bPhi}^\dagger 
  \, , \label{eq:phi_G_avg} 
\end{equation}
using the short-hand notation $\hat{\bG}_{\heta} \equiv  \hat{\bG}_V[\mbf{S}_{\heta}, \bS_{\heta}]$ and $\hat{\bPhi}_{\heta} \equiv \hat{\bPhi}_V[\mbf{S}_{\heta}, \bS_{\heta}]$, where $\mbf{S}_{\heta}$ and $\bS_{\heta}$ denote the self-energies for the disorder configuration $\heta$, see Eq.\ (\ref{eq:G_funct_eta}).
The corresponding average self-energies $\bar{\mbf{S}}$ and $\bar{\bS}$ are defined through the Dyson equations
\begin{equation}
  \bar{\mbf{S}}  = \bF - \bG_{\mbf{t} 0 0}^{-1} \bar{\bPhi} \, , \quad
  \bar{\bS}  = \bG_{\mbf{t} 0 0}^{-1} - \bar{\bG}^{-1} \label{eq:Sigm_P}
  \, ,
\end{equation}
where $\bG_{\mbf{t} 0 0}$ is the free propagator for the disorder-free system $\bG_{\mbf{t} 0 0} \equiv \bG_{\mbf{t} \heta 0}|_{\heta = \mbf{0}}$.
%
% and $\bar{\bPhi} $ and  $\bar{\bG} $ are the disorder-averaged condensate and connected Green's function, respectively, computed by evaluating Eqs.\ (\ref{eq:phi_funct_P}) and (\ref{eq:G_funct_P}) at the physical self-energies $\mbf{S}_{\bF \mbf{t} \heta V}$ and $\bS_{\bF \mbf{t} \heta V}$ for each disorder configuration $\heta$.
%
By insertion of the averaged Dyson equations [Eq.\ (\ref{eq:Sigm_P})] in the definitions of the averaged propagators $\bar{\bPhi}$ and $\bar{\bG}$ [Eq.\ (\ref{eq:phi_G_avg})] we obtain the relations
\begin{equation}
  \bar{\bPhi}
   =
  \Big\langle
      \mbf{A}_{\heta}^{-1}\mbf{B}_{\heta}
      \Big\rangle_P
  \, , \label{eq:AvgPhiFunc}
\end{equation}
\begin{equation}
  \bar{\bG} - \bar{\bPhi} \bar{\bPhi}^\dagger
  =
  \Big\langle
      [ \mbf{A}_{\heta} - \bS_{\heta} ]^{-1}
  -
     \left[ \mbf{A}_{\heta}^{-1}
       \mbf{B}_{\heta}
      \right]\left[\mbf{A}_{\heta}^{-1}
       \mbf{B}_{\heta}
      \right]^{\dagger}
  \Big\rangle_P
  \, . \label{eq:AvgGFunc}
\end{equation}
where
\begin{equation}
 \mbf{A}_{\heta}=\bar{\bG}^{-1} + \bar{\bS} - \heta \, , \quad
  \mbf{B}_{\heta}  =
  \bar{\mbf{S}} - \mbf{S}_{\heta} + [\bar{\bG}^{-1} + \bar{\bS}] \bar{\bPhi}.
  \end{equation}
The concomitant solution of Eq.\ (\ref{eq:AvgPhiFunc}) and (\ref{eq:AvgGFunc}) implicitly defines the two universal functionals
\begin{equation}
  \hat{\bar{\bPhi}}[ \bar{\mbf{S}}, \bar{\bS}, \{ \mbf{S}_{\heta}, \bS_{\heta} \}]
  = \bar{\bPhi}
  \, , \quad
  \hat{\bar{\bG}}[ \bar{\mbf{S}}, \bar{\bS}, \{ \mbf{S}_{\heta}, \bS_{\heta} \}]
  = \bar{\bG}
  \, .
  \label{eq:phi_funct_P}
  \label{eq:G_funct_P}
\end{equation}
%
% --------------------------------------------------------------------
%\textcolor{blue}{
% --------------------------------------------------------------------
%To describe the combined effect of disorder and interaction we rewrite $\ASE$ in terms of the universal averaged propagator functionals
%
%\begin{eqnarray}
%  \hat{\bar{\bPhi}}
%  & \equiv
%  \hat{\bar{\bPhi}}_{PV}\left[\{ \mbf{S}_{\heta}, \bS_{\heta} \} \right]
% \equiv
%  \left\langle \hat{\bPhi}_{\heta} \right\rangle_P
%  \, ,
%  %\label{eq:phi_funct_P}
%  \\
%  \hat{\bar{\bG}}
%  & \equiv
%  \hat{\bar{\bG}}_{PV}\left[\{ \mbf{S}_{\heta}, \bS_{\heta} \} \right]
%  \equiv
%  \left\langle \hat{\bG}_{\heta} - \hat{\bPhi}_{\heta} \hat{\bPhi}_{\heta}^\dagger \right\rangle_P
%  + \hat{\bar{\bPhi}} \hat{\bar{\bPhi}}^\dagger 
%  \, .
%  %\label{eq:G_funct_P} 
%\end{eqnarray}
%
%using the short-hand notation $\hat{\bG}_{\heta} \equiv  \hat{\bG}_V[\mbf{S}_{\heta}, \bS_{\heta}]$ and $\hat{\bPhi}_{\heta} \equiv \hat{\bPhi}_V[\mbf{S}_{\heta}, \bS_{\heta}]$ [see Eq.\ (\ref{eq:G_funct_eta})].
% --------------------------------------------------------------------
%} % blue
% --------------------------------------------------------------------
%
Using the universal averaged propagator functionals we define the extended averaged self-energy effective action
\begin{eqnarray}
  \ASE[\bar{\mbf{S}}, \bar{\bS}, \{ \mbf{S}_{\heta}, \bS_{\heta} \} ]
  =& \frac{1}{2} (\bF - \bar{\mbf{S}})^\dagger \bG_{\mbf{t} 0 0} (\bF - \bar{\mbf{S}})
  + \frac{1}{2} \Tr \ln [ - (\bG^{-1}_{\mbf{t} 0 0} - \bar{\bS})] \nonumber
  \\
  &+
  %\hat{\mathcal{T}}_{PV}
  \ALWSE
      [\bar{\mbf{S}}, \bar{\bS}, \{ \mbf{S}_{\heta}, \bS_{\heta} \} ]
  +
  \left\langle
  \LWSE[\mbf{S}_{\heta}, \bS_{\heta}]
  \right\rangle_P
  \, ,
  \label{eq:DisorderAveragedExtendedSFT}
\end{eqnarray}
where $\ALWSE$ is a universal functional of the averaged self-energies
\begin{eqnarray}
   \ALWSE \left[\bar{\mbf{S}}, \bar{\bS}, \{ \mbf{S}_{\heta}, \bS_{\heta} \} \right]
  \equiv
  - \frac{1}{2}
  \hat{\bar{\bPhi}}^\dagger
 \left ( \hat{\bar{\bG}}^{-1} + \bar{\bS} \right)
  \hat{\bar{\bPhi}}
  -
  \frac{1}{2}
  \left\langle
    \hat{\bPhi}_{\heta}^{\dagger}
   \left ( \hat{\bar{\bG}}^{-1} + \bar{\bS} - \heta \right)
    \hat{\bPhi}_{\heta}^{}
  \right\rangle_P
  \nonumber
  \\
   +
  \left\langle
  \left[ \bar{\mbf{S}}^{\dagger} - \mbf{S}^{\dagger}_{\heta}
    + \hat{\bar{\bPhi}}^\dagger \left( \hat{\bar{\bG}}^{-1} + \bar{\bS} \right) \right]
  \hat{\bPhi}_{\heta}
  \right\rangle_P
   +
  \frac{1}{2}
  \left\langle
  \Tr \ln \left[ -\left ( \hat{\bar{\bG}}^{-1} + \bar{\bS} - \heta - \bS_{\heta} \right) \right]
  \right\rangle_P
  \nonumber
  \\ -
  \frac{1}{2} \Tr \ln \left[ - \hat{\bar{\bG}}^{-1} \right]
  \, . \label{eq:ALWSE_Def}
\end{eqnarray}
Here, we have extended the variational space from the fixed-disorder self-energies $[\{ \mbf{S}_{\heta}, \bS_{\heta} \}]$ to both the fixed-disorder and average self-energies $[\bar{\mbf{S}}, \bar{\bS}, \{ \mbf{S}_{\heta}, \bS_{\heta} \} ]$. However, when evaluated at the physical self-energies, $\ASE$ takes the value of the disorder average of the self-energy functional $\langle \SE \rangle_P$, see \ref{App:T_deriv}, and is thus equal to the disorder averaged free energy $\Omega_{\bF \mbf{t} P V}$ by Eq.\ (\ref{eq:DisFreeEnergy_in_SE}).

The variations of $\ALWSE$ are derived in \ref{App:T_deriv} and give,   $\delta_{\bar{\mbf{S}}^{\dagger}} \ALWSE = \hat{\bar{\bPhi}}$ and $2 \delta_{\bar{\bS}} \ALWSE = \hat{\bar{\bG}}$, showing that $\ALWSE$ is the analogue of the $\LWSE$ functional for the averaged self-energies, as by Eq.\ (\ref{eq:vars_F}), $\delta_{\mbf{S}^{\dagger}_{\heta}} \LWSE = \hat{\bPhi}_{\heta}$ and $2 \delta_{\bS_{\heta}} \LWSE = \hat{\bG}_{\heta}$.
%
% --------------------------------------------------------------------
%\textcolor{blue}{
% --------------------------------------------------------------------
%\begin{eqnarray}
%   \ALWSE [\bar{\mbf{S}}, \bar{\bS}, \{ \mbf{S}_{\heta}, \bS_{\heta} \} ]
%  \equiv 
%%
%  - \frac{1}{2}
%  \hat{\bar{\bPhi}}^\dagger
%  ( \hat{\bar{\bG}}^{-1} + \bar{\bS} )
%  \hat{\bar{\bPhi}}+ \frac{1}{2}
%  \left\langle \hat{\bar{\bPhi}}_{\heta}^{\dagger}
%  ( \hat{\bar{\bG}}^{-1} + \bar{\bS} - \heta )^{-1} \hat{\bar{\bPhi}}_{\heta}^{}\right\rangle_P  \nonumber
%  \\
%   -
%  \frac{1}{2} \Tr \ln ( - \hat{\bar{\bG}}^{-1} )   + \frac{1}{2}
%  \left\langle
%  \Tr \ln [ - ( \hat{\bar{\bG}}^{-1} + \bar{\bS} - \heta - \bS_{\heta} ) ]
%  \right\rangle_P
%  \,  \label{eq:ALWSE_Def}.
%\end{eqnarray}
% --------------------------------------------------------------------
%} % blue
% --------------------------------------------------------------------
%
%Taking the variations of $\ALWSE$ with respect to $\bar{\mbf{S}}$ and $\bar{\bS}$ gives
%
%\begin{equation}
%  \delta_{\bar{\mbf{S}}} \ALWSE = \hat{\bar{\bPhi}}
%  \, , \quad
%  2 \delta_{\bar{\bS}} \ALWSE = \hat{\bar{\bG}}
%  \, ,\nonumber
%\end{equation}
%
%which shows that $\ALWSE$ is the analogue of the $\LWSE$ functional for the averaged self-energies, as by (\ref{eq:vars_F})
%
%\begin{equation}
% \ \delta_{\mbf{S}_{\heta}} \LWSE = \hat{\bPhi}_{\heta}
%  \, , \quad
%  2 \delta_{\bS_{\heta}} \LWSE = \hat{\bG}_{\heta}
%  \, .\nonumber
%\end{equation}
%
The functional derivatives with respect to the self-energies at fixed disorder configuration $\heta$ yield
\begin{eqnarray}
  \delta_{\mbf{S}^{\dagger}_{\heta}} \ALWSE = 
  - P(\heta) \hat{\bPhi}_{\heta}
  +
  ( \delta_{\mbf{S}^{\dagger}_{\heta}} \hat{\bPhi}_{\heta}^\dagger ) P(\heta)
  \mathcal{Q}_{\heta}
  \, , \\
  2 \delta_{\bS_{\heta}} \ALWSE = 
  - P(\heta) \hat{\bG}_{\heta}
  +
  2( \delta_{\bS_{\heta}} \hat{\bPhi}_{\heta}^\dagger ) P(\heta)
  \mathcal{Q}_{\heta}
  \,  ,
\end{eqnarray}
where $\mathcal{Q}_{\heta}$ is defined in \ref{App:T_deriv} and vanishes at the physical self-energies.

The variations of the averaged self-energy effective action $\ASE$ [Eq.\ (\ref{eq:DisorderAveragedExtendedSFT})] therefore give
\begin{eqnarray}
  \delta_{\bar{\mbf{S}}^{\dagger}} \ASE
  & =
  - \bG_{\mbf{t} 0 0}(\bF - \bar{\mbf{S}}) + \hat{\bar{\bPhi}},
  \\
  2 \delta_{\bar{\bS}} \ASE
  & =
  - [\bG_{\mbf{t} 0 0}^{-1} - \bar{\bS}]^{-1} + \hat{\bar{\bG}},
  \\
  \delta_{\mbf{S}^{\dagger}_{\heta}} \ASE
  & =
  - P(\heta) \hat{\bPhi}_{\heta}
  + \delta_{\mbf{S}^{\dagger}_{\heta}} \langle \LWSE[\mbf{S}_{\heta}, \bS_{\heta}] \rangle_P+
  ( \delta_{\mbf{S}^{\dagger}_{\heta}} \hat{\bPhi}_{\heta}^\dagger ) P(\heta)
  \mathcal{Q}_{\heta},
  \\
  2 \delta_{\bS_{\heta}} \ASE
  & =
  - P(\heta) \hat{\bG}_{\heta}
  + 2 \delta_{\bS_{\heta}} \langle \LWSE[\mbf{S}_{\heta}, \bS_{\heta}] \rangle_P+
  2( \delta_{\bS_{\heta}} \hat{\bPhi}_{\heta}^\dagger ) P(\heta)
  \mathcal{Q}_{\heta}
,
  \, .
\end{eqnarray}
Hence, at the physical self-energies
\begin{equation}
  \bar{\mbf{S}}, \bar{\bS}, \{ \mbf{S}_{\heta}, \bS_{\heta} \}
  =
  \bar{\mbf{S}}_{\bF \mbf{t} P V}, \bar{\bS}_{\bF \mbf{t} P V}, \{ \mbf{S}_{\bF\mbf{t} \heta V}, \bS_{\bF \mbf{t} \heta V} \}
  \, ,\nonumber
\end{equation}
the averaged self-energy effective action $\ASE$ is stationary
\begin{equation}
 \delta_{\mbf{S}^{\dagger}_{\heta}} \ASE =
 \delta_{\bS_{\heta}} \ASE =\delta_{\bar{\mbf{S}}^{\dagger}} \ASE =
 \delta_{\bar{\bS}} \ASE = 0,
 \label{eq:variat_princ}
\end{equation}
and equal to the average free energy
\begin{equation}
  \ASE [ \bar{\mbf{S}}_{\bF \mbf{t} P V}, \bar{\bS}_{\bF \mbf{t} P V}, \{ \mbf{S}_{\bF \mbf{t} \heta V}, \bS_{\bF \mbf{t} \heta V} \} ]
  = \Omega_{\bF \mbf{t} P V}
  \, .\nonumber
\end{equation}

The crucial part of the disorder-averaged self-energy effective action is that the functionals $\ALWSE$ and $\LWSE$ are universal, in the sense that they do not depend on the non-interacting propagator $\bG_{\mbf{t} 0 0}$ or the symmetry-breaking field $\bF$ (see \ref{App:Univ} for the explicit derivatives of $\ALWSE$ and Ref.\ \cite{BSFT} for the universality of $\LWSE$), but only on the interaction $V$, the disorder probability distribution $P(\heta)$, the disorder-dependent self-energies $\left\{\mbf{S}_{\heta},\bS_{\heta}\right\}$, and the average self-energies $\bar{\mbf{S}}$ and $\bar{\bS}$. In the following we will make use of this property in order to derive consistent approximations of $\ASE$.

% ----------------------------------------------------------------------
\subsection{Disorder-averaged self-energy functional theory}
\label{sec:Avg_SFT}
% ----------------------------------------------------------------------

A versatile approach to non-perturbative approximations of the self-energy effective action $\SE$ is the self-energy functional theory (SFT) pioneered by Potthoff \cite{Potthoff:2003aa, Potthoff:2003ab, Potthoff:2006aa, Springer:2012} for fermionic systems and later extended to bosonic systems \cite{Koller:2006aa,Arrigoni:2011aa,BSFT}. The formalism for systems with disorder has been developed for fermions in Ref.\ \cite{Pot_Dis} and applied in a variational cluster approximation (VCA) to bosons without symmetry breaking in Ref.\ \cite{Kanp_Disorder}. Here, we generalize the bosonic case to also include $U(1)$-symmetry-breaking and general reference systems.

We consider the general interacting bosonic system with quadratic disorder of Eq.\ (\ref{eq:H}), and introduce a second \emph{reference} system with the same interaction $\hat{V}$ and disorder $P(\heta)$ but with some arbitrary linear symmetry breaking field $\bF'$, arbitrary free propagator
\begin{equation}
  \bG_{\mbf{\Delta} \heta 0}^{-1}(\tau, \tau')
  =
    \delta(\tau - \tau')
  (- [\mbf{1} \otimes \sigma_z] \partial_{\tau'} - \heta)-\bDelta(\tau, \tau')
  \, , \nonumber
\end{equation}
and self-energy effective action $\ASEprime$. Here, the free propagator $ \bG_{\mbf{\Delta} \heta 0}$ is parametrized by replacing the hopping $\mbf{t}$ by a completely general matrix $\bDelta(\tau, \tau')$  \footnote{In e.g.\ the context of dynamical mean-field theory $\bDelta(\tau-\tau')$ would represent a retarded hybridization of an impurity with a non-interacting bath, while in the case of e.g.\ an instantaneous $\bDelta(\tau, \tau')=\mbf{t}'\delta(\tau-\tau')$ where $\mbf{t}'$ is diagonal in Nambu space, it can be considered to be a hopping amplitude.} .

Now, since the self-energy effective actions of both systems contain the same universal functionals $\hat{\mathcal{F}}_{PV}$ and $\ALWSE$ we can evaluate $\ASE$ in terms of $\ASEprime$ as
\begin{eqnarray}
  \ASE[\bar{\mbf{S}}, \bar{\bS}, \{ \mbf{S}_{ \heta}, \bS_{\heta} \}]
  =&
  \ASEprime[\bar{\mbf{S}}, \bar{\bS}, \{ \mbf{S}_{\heta}, \bS_{\heta} \}]+ \frac{1}{2} \Tr \ln
  \left[
   \bG_{\mbf{t}00}^{-1} - \bar{\bS} \right]  \nonumber
  \\ &-\frac{1}{2} \Tr \ln
  \left[ \bG_{\bDelta00}^{-1} - \bar{\bS} \right]+ \frac{1}{2} (\bF - \bar{\mbf{S}})^\dagger \bG_{\mbf{t} 0 0} (\bF - \bar{\mbf{S}}) \nonumber \\
    &  -\frac{1}{2} (\bF' - \bar{\mbf{S}})^\dagger \bG_{\bDelta 0 0} (\bF' - \bar{\mbf{S}}). 
\end{eqnarray}
The stationary condition in Eq.\ (\ref{eq:variat_princ}) now translates into
\begin{eqnarray}
  \frac{\delta \ASE }{ \delta \bar{\mbf{S}}^{\dagger} }
  & =
  \bG_{\bDelta 00}(\bF' - \bar{\mbf{S}})
  -
  \bG_{\mbf{t}00}(\bF - \bar{\mbf{S}} )=0
  \, , \label{eq:dSFTdS}
  \\
  2\frac{\delta \ASE }{ \delta \bar{\bS} }
  & =
  [\bG_{\bDelta00}^{-1} - \bar{\bS}]^{-1}
  -
  [\bG_{\mbf{t}00}^{-1} - \bar{\bS}]^{-1}=0
  \, . \label{eq:dSFTdSigma}  
\end{eqnarray}

If by an appropriate choice of $\bDelta$ and $\bF'$ the reference system can be made simple enough to be exactly solvable, one can go one step further and evaluate the original systems functional $\ASE$ at the physical self-energies of the reference system, i.e.\ at $\bar{\mbf{S}}_{\bF'\bDelta}\equiv\bar{\mbf{S}}_{\bF'\bDelta P V}$, $\bar{\bS}_{\bF'\bDelta}\equiv \bar{\bS}_{\bF'\bDelta P V}$,  and $\{\mbf{S}_{\bF'\bDelta \heta V},\bS_{\bF'\bDelta \heta V}\}$.
This produces the self-energy functional theory (SFT) approximation for the system and the SFT functional 
\begin{eqnarray}
  \SFT
  [\bar{\mbf{S}}_{\bF'\bDelta}, \bar{\bS}_{\bF'\bDelta}]
  =
  \Omega_{\bF'\bDelta PV}
  + \frac{1}{2} \Tr \ln
  \left[ \bG_{\mbf{t}00}^{-1} - \bar{\bS}_{\bF'\bDelta} \right] -\frac{1}{2} \Tr \ln
  \left[  \bG_{\bDelta 00}^{-1} - \bar{\bS}_{\bF'\bDelta}
  \right] \nonumber \\
 + \frac{1}{2} (\bF - \bar{\mbf{S}}_{\bF'\bDelta})^\dagger \bG_{\mbf{t} 0 0} (\bF - \bar{\mbf{S}}_{\bF'\bDelta})  -\frac{1}{2} (\bF' - \bar{\mbf{S}}_{\bF'\bDelta})^\dagger \bG_{\bDelta 0 0} (\bF' - \bar{\mbf{S}}_{\bF'\bDelta})
  \, , \label{eq:SFT_functional}
\end{eqnarray}
where we have used that $\ASEprime [\bar{\mbf{S}}_{\bF'\bDelta}, \bar{\bS}_{\bF'\bDelta},\{\mbf{S}_{\bF'\bDelta \heta V},\bS_{\bF'\bDelta \heta V}\}]= \Omega_{\bF'\bDelta PV}$, and 
\begin{equation}
  \SFT[ \bar{\mbf{S}}_{\bF'\bDelta}, \bar{\bS}_{\bF'\bDelta}] \equiv \ASE[ \bar{\mbf{S}}_{\bF'\bDelta}, \bar{\bS}_{\bF'\bDelta},\{\mbf{S}_{\bF'\bDelta \heta V},\bS_{\bF'\bDelta \heta V}\}]
  \, ,\nonumber
\end{equation}
is the self-energy effective action of the original system $\ASE$ restricted to the domain of physical self-energies of the reference system. Note that by replacing $\ASEprime [\bar{\mbf{S}}_{\bF'\bDelta}, \bar{\bS}_{\bF'\bDelta},\{\mbf{S}_{\bF'\bDelta \heta V},\bS_{\bF'\bDelta \heta V}\}]$ with the scalar $\Omega_{\bF'\bDelta PV}$, we eliminate all explicit dependencies on the fixed-disorder self-energies $\{\mbf{S}_{\bF'\bDelta \heta V},\bS_{\bF'\bDelta \heta V}\}$, such that the disorder-averaged self-energy effective action now only depends on the average self-energies  $\bar{\mbf{S}}_{\bF'\bDelta}$ and $\bar{\bS}_{\bF'\bDelta}$. 

The domain of $\SFT$ is therefore defined by the average physical self-energies of the reference system  ($\bar{\mbf{S}}_{\bF'\bDelta}$ and $ \bar{\bS}_{\bF'\bDelta}$) and parametrized by $\bDelta$ and $\bF'$.
By generalizing the variational principle of Eq.\ (\ref{eq:variat_princ}) to the restricted domain we obtain a thermodynamically optimal approximation when the self-energy variations are zero on the domain, i.e. we seek $\bDelta$ and $\bF'$ such that
\begin{equation}
  \frac{\delta \SFT }{ \delta\bar{\mbf{S}}^{\dagger}_{\bF'\bDelta} }=  \frac{\delta \SFT }{ \delta \bar{\bS}_{\bF'\bDelta} }=0,\nonumber
\end{equation}
which by
\begin{eqnarray}
  \frac{\delta \SFT }{ \delta \bar{\mbf{S}}^{\dagger}_{\bF'\bDelta} }
  & = \frac{\delta \SFT }{ \delta \bF' }\left[ \frac{ \delta \bar{\mbf{S}}^{\dagger}_{\bF'\bDelta} }{\delta \bF' }\right]^{-1}+\frac{\delta \SFT }{ \delta \bDelta } \left[ \frac{ \delta \bar{\mbf{S}}^{\dagger}_{\bF'\bDelta} }{\delta \bDelta }\right]^{-1}
  \, , 
  \\
  \frac{\delta \SFT }{ \delta \bar{\bS}_{\bF'\bDelta} }
  & =
 \frac{\delta \SFT }{ \delta \bF' }\left[  \frac{  \delta \bar{\bS}_{\bF'\bDelta} }{\delta \bF' }\right]^{-1}+\frac{\delta \SFT }{ \delta \bDelta }\left[  \frac{\delta \bar{\bS}_{\bF'\bDelta} }{\delta \bDelta }\right]^{-1}
  \, , 
\end{eqnarray}
can be fulfilled if
\begin{equation}
\frac{\delta \SFT }{ \delta \bF' }=\frac{\delta \SFT }{ \delta \bDelta } =0. \label{eq:stat_FD}
\end{equation}

By the SFT approximation, the entire complexity of the original lattice system has therefore been reduced to finding stationary solutions of the functional in Eq.\ (\ref{eq:SFT_functional}) in terms of the variational parameters $\bF' $ and $\bDelta$.

Note that by restricting $\bDelta$ to be local, but keeping its full imaginary-time dependence, SFT reduces to a disorder-averaged version of bosonic dynamical mean-field theory \cite{Byc_BDMFT, Hu:2009qf, Hubener:2009cr, BDMFT, BDMFT1}, see \ref{sec:Dis_BDMFT}.

% ----------------------------------------------------------------------
% ----------------------------------------------------------------------
% ----------------------------------------------------------------------

% ----------------------------------------------------------------------
\section{Disordered Bose-Hubbard model}
\label{sec:model}
% ----------------------------------------------------------------------

As a simple application of the formalism derived in Sec.\ \ref{sec:theory} we study the disordered Bose-Hubbard model (BHm) on the cubic lattice with uncorrelated box disorder.
The Hamiltonian has the form
\begin{equation}
  H=
  -J\sum_{\langle i,j\rangle}b^{\dagger}_{i}b^{}_{j}
  +\frac{U}{2}\sum_{i}\bc_i\bc_i\ba_i\ba_i
  +\sum_i\left(\eta_i \!-\! \mu\right)\n_i,
  \nonumber
\end{equation}
where $b_i^{\left(\dagger\right)}$ creates (annihilates) a boson at site $i$, $\n_i=b^\dagger_i b_i$ is the occupation number operator, ${\langle i,j\rangle}$ denotes summation over nearest neighbors, $J$ is the hopping amplitude, $U$ the on-site interaction, and $\mu$ the chemical potential. 
The local disorder potentials $\eta_i$ are uncorrelated and are assumed to have a flat probability distribution
\begin{equation}
  P(\heta)=\prod_i p(\eta_i)
  \, , \, \,
  p(\eta)=
  \left\lbrace \begin{array}{ll}
    1/(2D) & \rm{, if }\left|\eta\right|\leq D \\
    0 & \rm{, else}
  \end{array}
  \right.
  \label{eq:boxed_disorder}
\end{equation}
where $D$ is the disorder strength.
Thus, the free propagator [Eq.\ (\ref{eq:SingleParticlePropagator})] is given by
\begin{equation}
  \bG_{\mbf{t} \heta 0}^{-1}(\tau, \tau')
   =
   \delta(\tau - \tau')
   \left(
     -[\mbf{1} \otimes \sigma_z] \partial_{\tau'} +
     \left[\left(J\delta_{\langle i,j\rangle} + [\mu - \eta_i]\delta_{ij}\right) \otimes \mbf{1}\right]
   \right)
  \, ,
\end{equation}
where $\delta_{\langle ij\rangle}$ is non-zero only for nearest neighbors $\langle i,j\rangle$.

In addition to the superfluid and Mott insulating phase of the clean (i.e.\ non-disordered) model, the groundstate phase diagram of the disordered BHm exhibits the Bose glass phase.
It is an insulating compressible phase that always intervenes between the superfluid and the Mott insulator at finite disorder ($D > 0$) \cite{Gur_09}.
The phase is composed of local regions, including both strongly localized atomic levels, and isolated superfluid lakes which locally close the many-body gap. Since these superfluid lakes are spatially separated, global phase coherence is not reached, yielding a zero superfluid response. Hence, while the compressibility of the Bose glass is finite, the global condensate order parameter is zero as is the many-body gap. In our SFT formalism we therefore distinguish the Bose glass and the superfluid phase by the disorder-averaged condensate $\bar{\bPhi}$. In the mean-field approach of Ref.\ \cite{Rieger_13} a different criterion was used, analyzing the spatial percolation of superfluid regions (i.e. isolated regions with non-zero quasi-condensates), allowing for a zero global superfluid response even though $\bar{\bPhi}\neq 0$. As in this work we will treat only disorder-averaged translational-invariant quantities, such a real-space percolation of the condensate is not analyzed directly.

% ----------------------------------------------------------------------
%\subsection{SFA3 reference system}
\subsection{Minimal reference system}
\label{sec:SFA3}
% ----------------------------------------------------------------------

In this first application of SFT with symmetry breaking to the disordered BHm, we will make use of the simplest possible reference system, comprising a single bosonic mode per site.
In this case the reference system Hamiltonian in the thermodynamic limit reads
\begin{equation}
H'_{\heta}\left[\bF',\bDelta\right]=\sum_{i}\tilde{H}'_{i, \eta_i}\left[\bF',\bDelta\right],\nonumber
\end{equation}
where the sum runs over an infinite number of independent single-site Hamiltonians
\begin{equation}
  \tilde{H}'_{i, \eta_i}
   =
   \frac{U}{2}\bc_i \bc_i \ba_i \ba_i +\left(\eta_i \!-\! \mu\right)\n_i
   + \bbc_i \bF' + \bbc_i \frac{\bDelta}{2} \bba_i
   . \label{H_i}
\end{equation}
The reference system is parametrized by three real translationally-invariant parameters $F'$, $\Delta_{00}$, and $\Delta_{01}$, where $\bF'=\left(F',F'\right)$ and
$\bDelta(\tau-\tau')=\delta(\tau-\tau')\bDelta$ is instantaneous in imaginary time and site-local as
\begin{equation}
  \bDelta(\tau-\tau') =
  \delta(\tau-\tau')
  [\delta_{ij}
  \otimes
  (\Delta_{00} \mbf{1} + \Delta_{01} \sigma_x)]
  \, .\nonumber
\end{equation}
This minimal reference system yields a non-perturbative self-energy functional approximation that we denote by SFA3. It has previously been shown to yield quantitatively correct results for the clean BHm, comparing with numerically exact QMC results \cite{BSFT}.

In the case of uncorrelated disorder considered here, the disorder-averaging of observables [Eq.\ (\ref{eq:EnsembleAverage})] gives translationally invariant results, see \ref{sec:uncorrelated}.
The disorder-averaged free energy of the SFA3 reference system is therefore given by
\begin{equation}
\Omega_{\bF' \bDelta P U} = N \langle \Omega_{i,\bF' \bDelta \eta_i U} \rangle_p\nonumber
\end{equation}
where $N$ is the number of lattice sites, $\Omega_{i,\bF' \bDelta \eta_i U}$ is the free energy of a single site in the reference system, and $\langle f(\eta) \rangle_p \equiv \int d\eta\, p(\eta) f(\eta)$. Analogously, the propagators are obtained as
\begin{equation}
  \bar{\bG}_{\bF' \bDelta P U}(\tau-\tau')
   =
  \delta_{ij}
  \otimes
  \langle
  \bG_{ii, \bF' \bDelta \eta_i U}(\tau-\tau')
  \rangle_p
  \, , \quad
  {\bar{\bPhi}}_{\bF' \bDelta P U}
   =
  \langle
  {\bPhi}_{i, \bF' \bDelta \eta_i U}
  \rangle_p
  \, .
\end{equation}
Hence, to evaluate the disorder-averaged quantities of the reference system it suffices to solve the single-site Hamiltonian of Eq.\ (\ref{H_i}) for all possible values of $\eta_i$ and then average the result over the probability distribution $p(\eta)$. The corresponding average self-energies $\bar{\mbf{S}}_{\bF'\bDelta PU}$ and $\bar{\bS}_{\bF'\bDelta PU}$ of the reference system are then obtained from Eq.\  (\ref{eq:Sigm_P}).

Physical solutions of the lattice system can be found by searching for stationary values of the SFT functional in Eq.\ (\ref{eq:SFT_functional}) fulfilling Eq.\ (\ref{eq:stat_FD}) using a standard root solver to find the point with zero gradient. This procedure is identical to the algorithm detailed in Ref.\ \cite{BSFT}.
Once a stationary solution is found, lattice quantities can be computed using the corresponding self-energies at stationarity as detailed in \ref{sec:Dis_Lattice}.

% ----------------------------------------------------------------------
\subsection{Atomic limit}
\label{sec:atomic_limit}
% ----------------------------------------------------------------------

As in this work we mainly analyze the behavior of the disordered BHm at large interactions $U/J\gg 1$, we want to compare to the analytic atomic limit of having decoupled sites, i.e.  $J=0$. In this section we analyze the properties of the infinite system in this limit. 

% ----------------------------------------------------------------------
\subsubsection{Local occupations}
\label{Local_Occ}

We start by analyzing the local occupations as a function of disorder in the atomic limit. For $J=0$, we can have a local occupation $n_i=\langle \n_i \rangle$ at zero temperature if the local potential $\eta_i$ takes values $\eta_{\rm{min}}(n_i)<\eta_i<\eta_{\rm{max}}(n_i)$. In order to derive this, we turn to the local energy of the decoupled site $i$ with occupation number $n_i$ and local potential $\eta_i$, i.e.\
\begin{equation}
E_{\rm SS}(n_i,\eta_i)=\frac{U}{2}n_i\left(n_i-1\right)+\left(\eta_i-\mu\right)n_i,
\nonumber
\end{equation} 
The groundstate will have local occupation larger than $n_i-1$ if $E_{\rm SS}(n_i,\eta_i)<E_{\rm SS}(n_i-1,\eta_i)$, i.e. if
\begin{equation}
\eta_i<\eta_{\rm{max}}(n_i)= \left\lbrace \begin{array}{ll} 
\mu-U(n_i-1) & \textrm{if $n_i>0$}, \\
\infty &  \textrm{if $n_i=0$}, \end{array} \right.
\label{Eta_min}
\end{equation} 
where we used that the local occupation $n_i$ is bounded from below by zero, and therefore $\eta_{\rm min}(0)=\infty$.
Additionally, in order to have a local occupation of $n_i$ we need to fulfill the condition $E_{\rm SS}(n_i,\eta_i)< E_{\rm SS}(n_i+1,\eta_i)$, resulting in
\begin{equation}
\eta_i>\eta_{\rm min}(n_i)=\mu-Un_i.
\label{Eta_max}
\end{equation} 

As the minimum possible value of $\eta_i$ is $-D$ [see Eq.\ (\ref{eq:boxed_disorder})], this implies that the maximal possible local occupation $n_{\rm max}$ is given by
\begin{equation}
n_{\rm max}=\left\lfloor \frac{D+\mu}{U}+1\right\rfloor.
\label{n_max}
\end{equation} 
Furthermore, as the maximum value of $\eta_i$ is $D$ and the local occupation is bounded from below by $n_i=0$, we have a minimal possible local occupation of
\begin{equation}
n_{\rm min}= {\rm Max}\left\lbrace\left\lceil \frac{\mu-D}{U}\right\rceil,0\right\rbrace.
\label{n_min}
\end{equation} 

We can use the information above to derive the probability of sites with occupation $n$ in the infinite system. We denote this quantity by $r_n$, defined as the number of sites with local occupation $n$ divided by the total number of sites, which can be computed by
 \begin{equation}
r_n=\left\lbrace \begin{array}{l} 0 \textrm{   if $n<n_{\rm min}$ or $n>n_{\rm max}$},\\ 
 \frac{1}{2D}\left({\rm Min}\left\lbrace D,\eta_{\rm max}(n)\right\rbrace-{\rm Max}\left\lbrace-D,\eta_{\rm min}(n)\right\rbrace\right)  \textrm{   else}. \end{array}\right.
\nonumber
\end{equation} 
We can use the probabilities $r_n$ in order to derive expressions for the total density $n$ and the interaction energy $E_{\rm int}$, given by
 \begin{equation}
n=\sum_{m=0}^{\infty} r_m m,
\quad
E_{\rm int}=\frac{U}{2}\sum_{m=0}^{\infty} r_m \left(m^2-m\right).
\end{equation} 

Note that, while the values of $r_n$ depend on the disorder distribution $P(\heta)$, the values of $D$ where they become non-zero [and therefore the maximal and minimal possible occupations for a given disorder strength in Eqs.\ (\ref{n_max}) and (\ref{n_min})] depend only on the maximal and minimal values of the local potential $\pm D$ (and the global parameters $\mu$ and $U$). These are therefore universal, in the sense that they do not depend on the disorder distribution $P(\heta)$ as long as it is uncorrelated and bounded [i.e.\ with $p\left(\left|\eta_i\right|>D\right)=0$]. 

% ----------------------------------------------------------------------
\subsubsection{Local excitations}
\label{Loc_Ex}
% ----------------------------------------------------------------------

%
On a single-site level, the process $(n_i\rightarrow n_i+1)$ on site $i$ leads to the energy difference
\begin{equation}
\Delta E\left[n_i\rightarrow n_i+1\right]=Un_i-\mu+\eta_i.
\label{DEp}
\end{equation}
As discussed in Sec.\ \ref{Local_Occ}, in the groundstate we can have a local occupation of $n_i$ only if $\eta_{\rm min}(n_i)\leq\eta_{i}\leq\eta_{\rm max}(n_i)$, see Eqs.\ (\ref{Eta_min}) and (\ref{Eta_max}). If we average over all sites $i$, we therefore find, that the local processes $(n_l\rightarrow n_l+1)$, with local groundstate occupations $n_l$, span over the energy range given by
\begin{equation}
{\rm Max}\left\lbrace 0, Un_l-\mu-D  \right\rbrace\leq\Delta E\left[n_l\rightarrow n_l+1\right]\leq  \left\lbrace \begin{array}{l} 
D-\mu \textrm{ if $n_l=0$}, \\
{\rm Min}\left\lbrace U, Un_l-\mu+D  \right\rbrace \textrm{ else}. \end{array} \right.
\label{DEp1}
\end{equation}
We can further derive the energy difference for the opposite process in the same way, yielding
\begin{equation}
 {\rm Max}\left\lbrace 0,\mu-D -Un_l  \right\rbrace\leq\Delta E\left[n_l+1\rightarrow n_l\right]\leq  
{\rm Min}\left\lbrace U, \mu+D-Un_l  \right\rbrace.
\label{DEm1}
\end{equation}

The disorder-averaged local spectral function is defined as
\begin{equation}
A_{\rm loc}(\omega)=-\frac{1}{N\pi}\sum_{i}{\rm Im}\left[\bar{G}_{ii}(\omega)\right].
\nonumber
\end{equation}
At zero temperature we have
\begin{equation}
G_{ii,\heta}(\omega) =\sum_{n\neq  {\rm GS}}\left(\frac{\left|\left\langle n \left| b^{\dagger}_i \right| {\rm GS} \right\rangle \right|^{2}}{E_{\rm GS} -E_{n}+\omega^{+}} +\frac{\left|\left\langle n \left| b^{\phantom\dagger}_i \right|  {\rm GS} \right\rangle \right|^{2}}{E_{\rm GS}-E_{n}-\omega^{+}}\right), \nonumber 
%\label{Gloc}
\end{equation}
where GS is the groundstate, the sum runs over all other eigenstates, $E_{n}$ is the energy of eigenstate $n$, and $\omega^+=\omega+i\epsilon$ with a small broadening parameter $\epsilon$. Disorder-averaging over an infinite number of configurations therefore yields a translational invariant local Green's function
\begin{eqnarray}
\bar{G}_{ii}(\omega) =& \int_{-D}^{D} d\eta p(\eta)\left(\frac{\tilde{n}(\eta)+1}{\omega^{+}-\Delta E\left[\tilde{n}(\eta)\rightarrow \tilde{n}(\eta)+1\right]}\right. \nonumber \\
&+\left.\frac{\tilde{n}(\eta)}{\Delta E\left[\tilde{n}(\eta)-1\rightarrow \tilde{n}(\eta)\right]-\omega^{+}}\right) , 
\end{eqnarray}
where
\begin{equation}
\tilde{n}\left(\eta_{\rm{min}}(n)<\eta<\eta_{\rm{max}}(n)\right)=n. \nonumber
\end{equation}

Using Eqs.\ (\ref{DEp1}) and (\ref{DEm1}) we therefore find that the resonances of the spectral function for the processes $(n\rightarrow n+1)$ are bounded by Eq.\ (\ref{DEp1}), while the processes $(n+1\rightarrow n)$  are bounded by Eq.\ (\ref{DEm1}). Therefore, the effect of the disorder strength $D$ on the spectral function in the atomic limit -- which in the absence of disorder consists of sharp delta peaks -- is to broaden the peaks to a width which is proportional to $D$.
 A consequence of this is that, apart from the process $(0\rightarrow 1)$, all local resonances are bounded by $-U\leq\omega\leq U$. Further, it can easily be shown that for $D\geq mU/2$, with integer $m$, we have $\omega_{\rm max}(n\rightarrow n+1)>\omega_{\rm min}(n+m\rightarrow n+m+1)$, leading to an overlap of the processes $(n\rightarrow n+1)$ and $(n+m\rightarrow  n+m+1)$ (and equivalently for the reversed particle-removal processes).
\begin{figure}
\centering
\includegraphics[width=300pt]{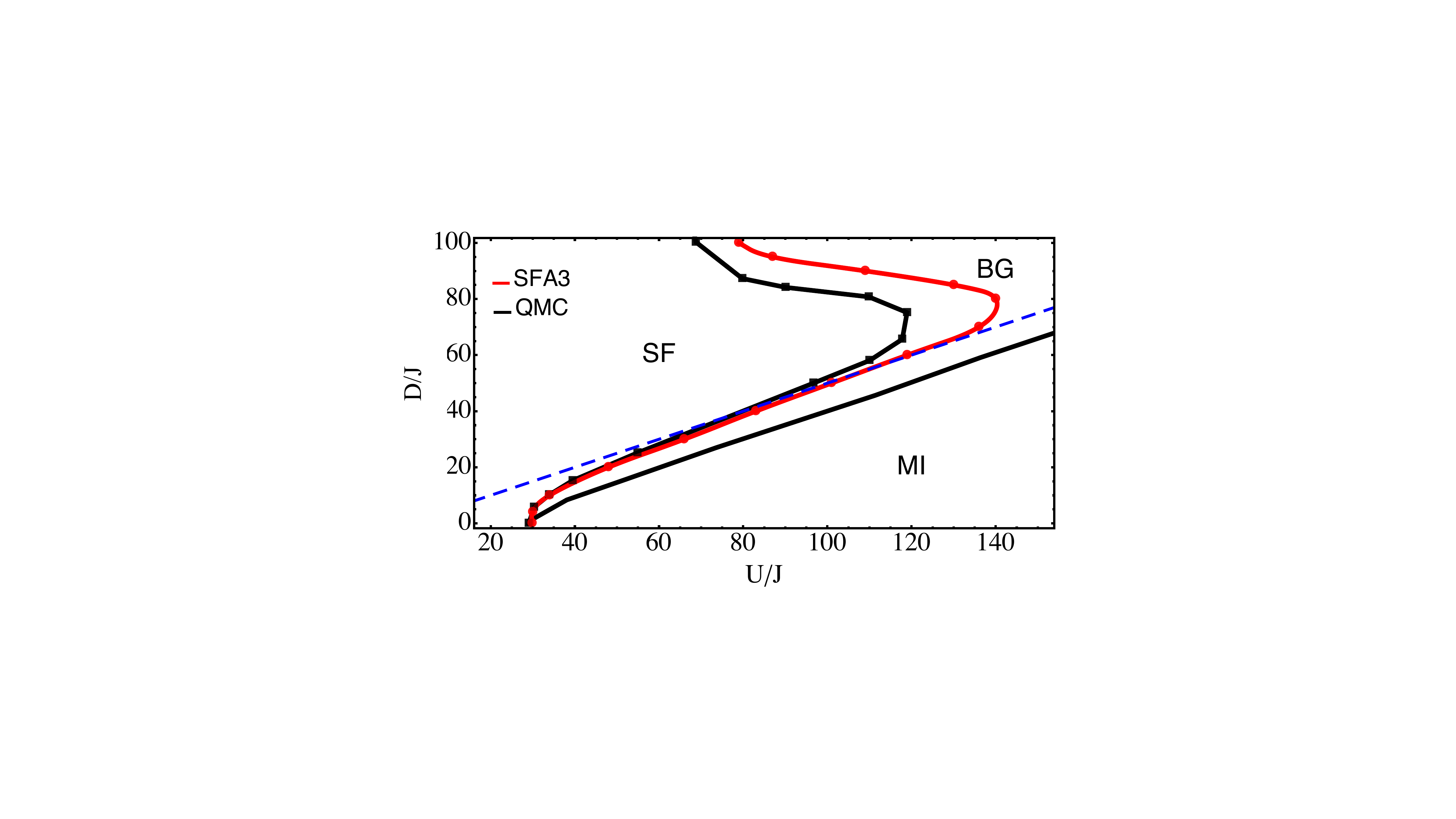}
  \caption{\label{fig:pha_da} Groundstate phase diagram of the disordered BHm with box disorder at fixed density $n=1$. The SFA3 results on the superfluid to Bose glass transition are shown in red, while the QMC results (black squares) are taken from Ref.\ \cite{Gur_09}. The blue dashed line indicates the point where doubly occupied sites are activated in the atomic limit.}
\end{figure}

% --------------------------------------------------------------------
\section{Results}
\label{sec:results}
% --------------------------------------------------------------------

We analyze the BHm with box disorder using SFT with an SFA3 reference system, see Sec.\ \ref{sec:SFA3}. The calculations are compared to disorder-averaged path integral quantum Monte Carlo (QMC) \cite{Rev_Lode,QMC_Cap,Gur_09} simulations on a finite cubic lattice of $8^3$ sites. In the strongly-interacting case we further compare to analytic results in the atomic limit (i.e.\ the limit of zero hopping $J=0$), detailed in Sec.\ \ref{sec:atomic_limit}. 
The resulting groundstate phase diagram computed with SFA3 at large interactions for fixed density $n=1$ is shown in Fig.\ \ref{fig:pha_da} together with the QMC results of Ref.\ \cite{Gur_09}.
%
%In Fig.\ \ref{fig:pha_da} we show the resulting groundstate phase diagram computed with SFA3 at large interactions for fixed density $n=1$, comparing to QMC results of Ref.\ \cite{Gur_09}.
%
The groundstate phases exhibited by the system are the superfluid, the Mott insulator, and the Bose glass.
For the ordered BHm ($\heta = \mbf{0}$), the SFA3 approximation showed remarkable agreement with exact QMC results \cite{BSFT}. The phase diagram in Fig.\ \ref{fig:pha_da} shows that this remains true also for weak disorder $D/J\lesssim 30$, where the SFA3 superfluid to Bose glass transition line shows excellent agreement with the QMC result.
%
%As expected from the high accuracy of SFA3 calculations in the clean BHm \cite{BSFT}, for relatively low disorder ($D/J\lessapprox 30$) the SFA3 superfluid to Bose glass transition line shows excellent agreement with the QMC data.

For stronger disorder the situation changes, in particular in the so-called \emph{superfluid finger}, i.e., the narrow region of the superfluid phase extending to large values of the relative interaction strength $U/J$.
In the finger, the condensate density $\rho_c=\frac{1}{2}{\bar{\Phi}}^{\dagger}\bar{\Phi}$ is extremely low, and therefore very hard to resolve experimentally \cite{Gur_09}.
Small deviations from numerically exact results in the SFA3 calculations therefore lead to a notable shift in the phase boundaries and an overestimation of the extent of the superfluid finger, as seen in Fig.\ \ref{fig:pha_da}.
At even larger disorder when leaving the superfluid finger, the discrepancy between SFA3 and QMC results is reduced.
The Mott insulator to Bose glass transition at fixed density $n=1$ is very hard to resolve numerically (unlike the transition at fixed chemical potential discussed later), as the finite compressibility in the Bose glass close to the phase boundary is exponentially small \cite{Gur_09}. Instead, in Fig.\ \ref{fig:pha_da} we show analytic results on the phase boundary from Ref.\ \cite{Gur_09}. 

Note that, while it is always possible to find a Bose glass stationary solution at strong disorder in mean-field approaches (by setting the condensate to zero), arithmetically averaged mean-field always finds a groundstate with a finite condensate order parameter and lower free energy \cite{Arithm_Dis,Rieger_13}. In the context of SFT, the mean-field approximation can be understood as neglecting the kinetic contributions of uncondensed bosons in the self-energy functional \cite{BSFT}. By including these contributions in our SFA3 calculations, we are able to change the energy balance with respect to the mean-field approach, yielding a phase transition to the uncondensed Bose glass.

% --------------------------------------------------------------------
\subsection{Strongly-interacting Bose glass phase}
\label{sec:strong_BG}
% --------------------------------------------------------------------

Using the local occupation probabilities $r_n$ of Sec.\ \ref{Local_Occ}, it is possible to distinguish different regimes of the Bose glass in the atomic limit: the qualitative behavior of the Bose glass will change every time the disorder strength is large enough to activate a given local occupation $n$ (i.e.\ if the probability of finding a site with local occupation $n$, $r_n$, becomes non-zero as a function of $D$).
Coming from the Mott-insulating groundstate at density $n=1$ (where $r_{n\neq 1}= 0$) and increasing the disorder strength $D$, as we enter the Bose glass one of the probabilities $r_0$ and $r_2$ becomes non-zero, as either empty or doubly-occupied sites are activated by the disorder depending on the value of the chemical potential. When the disorder is increased further, also higher occupancies are activated and other probabilities $r_n$ become non-zero every time we enter a new regime of the strongly-interacting Bose glass.

\begin{figure}
\centering
\includegraphics[width=300pt]{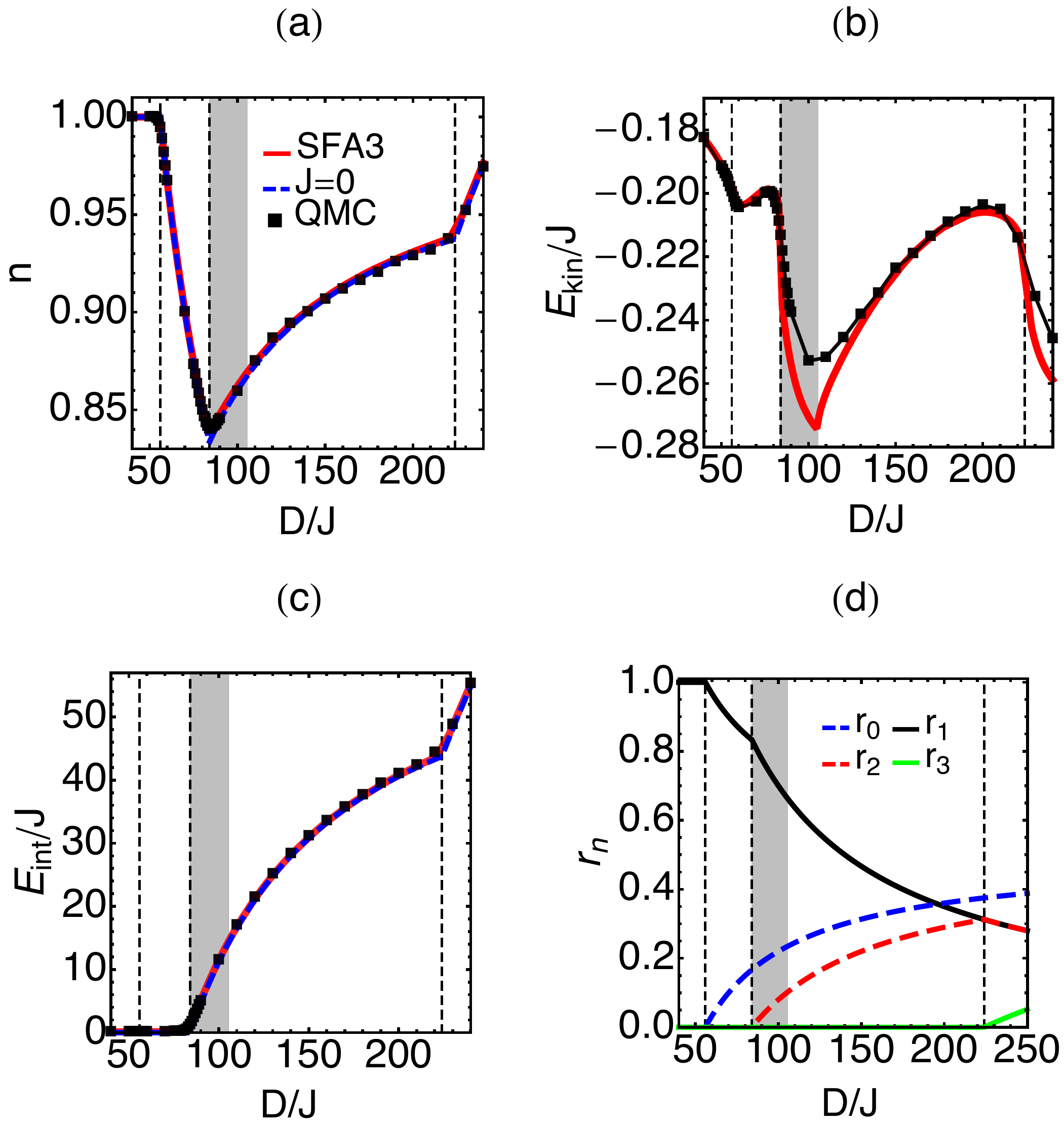}
  \caption{\label{fig:bg_swe} Observables of the disordered BHm as a function of disorder strength $D/J$ for $U/J=140$, $\mu/U=0.4$ and $T/J=0.1$. (a) Density $n$ computed with SFA3 (red), QMC (black squares), and in the atomic limit ($J=0$, blue dashed). (b) Kinetic energy per site $E_{\rm{kin}}/J$ computed with SFA3 (red), and QMC (black squares). (c) Interaction energy per site $E_{\rm{int}}/J$ computed with SFA3 (red), QMC (black squares), and in the atomic limit ($J=0$, blue dashed). (d) Probabilities of having sites with local occupation $0$ ($r_0$, blue dashed), $1$ ($r_1$, black), $2$ ($r_2$, red dashed), and $3$ ($r_3$, green), as computed in the atomic limit. The vertical dashed lines show the transitions between the different regimes in the atomic limit, while the grey area is where the non-local Green's function of SFA3 develops a pole, indicating the presence of isolated quasi-condensates.}
\end{figure}

\begin{figure}
\centering
\includegraphics[width=\columnwidth]{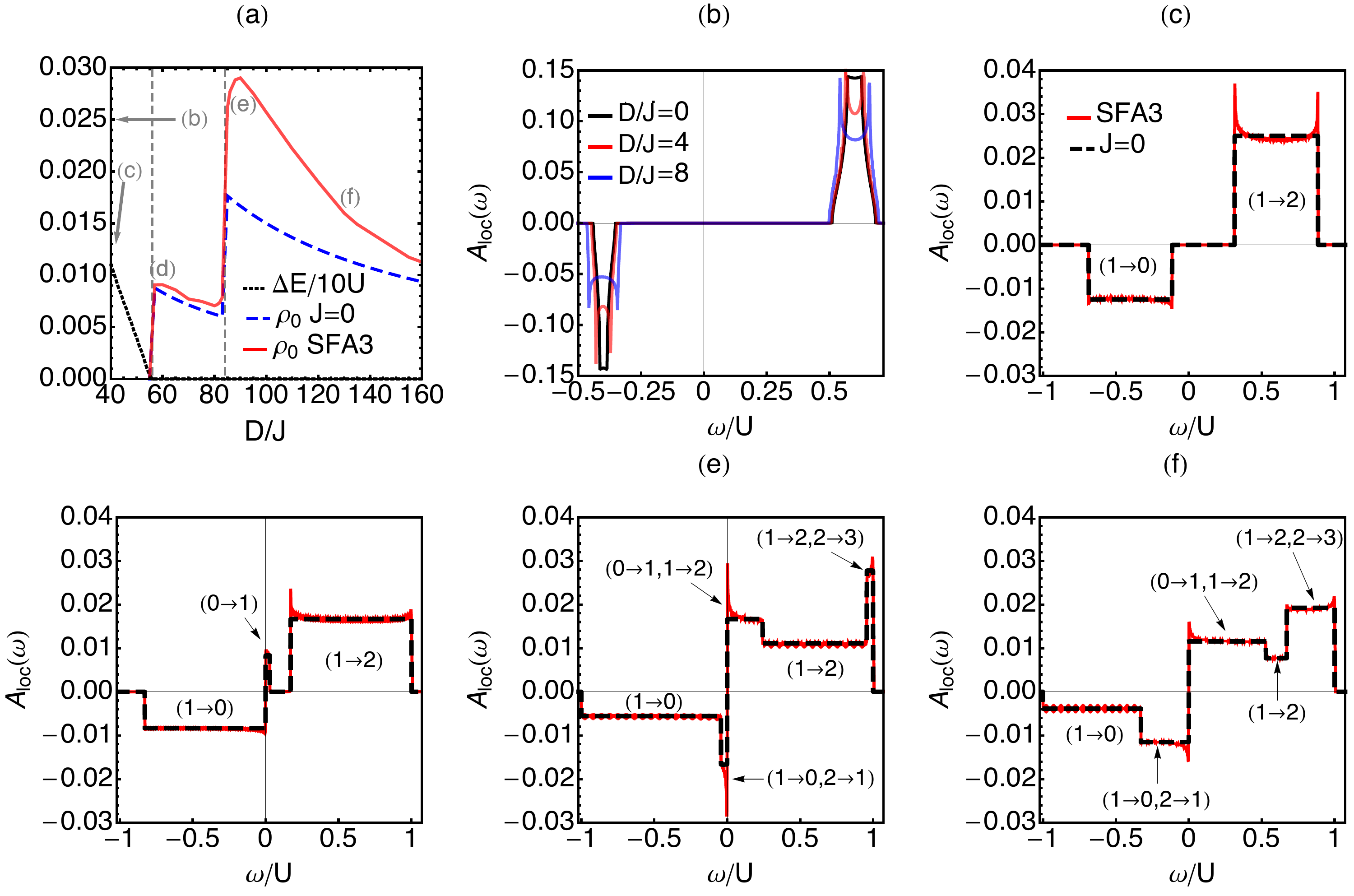}
  \caption{\label{fig:bg_spec} Properties of the local spectral function of the disordered BHm as a function of disorder strength $D/J$ for $U/J=140$, $\mu/U=0.4$ and $T/J=0.1$. (a) Mott gap [black dotted, renormalized as $\Delta E/(10U)$ for plotting purposes], and spectral weight around $\omega=0$, $\rho_0$ [see Eq.\ (\ref{eq:rho_0})] computed with SFA3 (red) and in the atomic limit (blue dashed). (b) Local spectral function for $D/J=0$ (black), $D/J=4$ (red), and $D/J=8$ (blue). (c-f) Local spectral functions computed with SFA3 (red) and in the atomic limit (black dashed) for $D/J=40$ (c), $D/J=60$ (d), $D/J=90$ (e), and $D/J=130$ (f). The involved transitions from local occupation $x$ to local occupation $y$ are denoted as $(x\rightarrow y)$.}
\end{figure}

While the atomic limit shows sharp transitions between the different regimes (see the values of $r_n$ in Fig.\ \ref{fig:bg_swe}d), for finite hopping $J$, the kinetic fluctuations turn the transitions into crossovers. However, as we will discuss in this section, in the case of strong interactions the qualitative behavior of local observables changes drastically also in our numerical results whenever a new regime is entered. For the sweep in disorder strength of Fig.\ \ref{fig:bg_swe} the results for local quantities such as the density (Fig.\ \ref{fig:bg_swe}a) and the interaction energy (Fig.\ \ref{fig:bg_swe}c) show perfect agreement between the analytic results in the atomic limit and both SFA3 and QMC results, except right at the transition/crossover between the different regimes. In fact, the kinetic energy (Fig.\ \ref{fig:bg_swe}b) -- which is the dominating additional contribution of the finite hopping in SFA3 and QMC, as compared to the atomic limit -- is orders of magnitude smaller than the interaction energy at large disorder. In the following we will discuss these different strongly-interacting regimes in more detail, analyzing the qualitative behavior of the observables in Fig.\ \ref{fig:bg_swe} and extracting additional information from the corresponding local spectral functions $A_{\rm{loc}}(\omega)$ shown in Fig.\ \ref{fig:bg_spec}.

We start at $D=0$, i.e., in the non-disordered Mott insulator. As every site has the same local occupation $n_i=1$, the local spectral function (Fig.\ \ref{fig:bg_spec}b) is characterized by the two Hubbard bands corresponding to the transitions $(1\rightarrow 0)$ at negative frequencies and $(1\rightarrow 2)$ at positive frequencies. While in the atomic limit these resonances would correspond to delta-peaks, at finite hopping the shape of the spectral function depends on the non-interacting dispersion and its bandwidth $W=2zJ$, where $z=6$ is the coordination number of the lattice.
In particular, the unit filling Mott insulator lower and upper Hubbard bands have the bandwidths $W$ and $2W$ respectively, see Ref.\ \cite{Strand:2015ac} for a derivation.
For weak disorder $D < W$ the qualitative behavior remains the same. However, the Hubbard bands are broadened by the finite disorder strength $D$ and the spectral weight at the center of the bands is reduced, see Fig.\ \ref{fig:bg_spec}b.

The situation changes when $D>W$ (see Fig.\ \ref{fig:bg_spec}c), where the spectral function is more similar to the one predicted by the atomic limit: as discussed in Sec.\ \ref{Loc_Ex}, the width of the Hubbard bands now is fully determined by the disorder strength $D$, and the dispersive features of the spectral function are lost. As we are still in the Mott phase, the spectral function shows a finite gap, defined as the minimal distance between the Hubbard bands and $\omega=0$. As the disorder strength $D$ is increased, so does the kinetic energy (see Fig.\ \ref{fig:bg_swe}b), due to increasing kinetic fluctuations, while the gap of the spectral function decreases (see Fig.\ \ref{fig:bg_spec}a). 

At $D\approx \mu$ ($D/J\approx 56$) the gap goes to zero, and we enter the Bose glass phase. The disorder activates empty sites (i.e. $r_0>0$, see Fig.\ \ref{fig:bg_swe}d), and as a consequence the density drops (Fig.\ \ref{fig:bg_swe}a), while the kinetic energy decreases (Fig.\ \ref{fig:bg_swe}b). The lower Hubbard band now extends to $\omega=0$, and we find a finite spectral weight at small positive frequencies corresponding to the local excitation $(0 \rightarrow 1)$ of the unoccupied sites (Fig.\ \ref{fig:bg_spec}d).
In order to study trends in the spectral weight at zero frequency $\omega=0$, we introduce the spectral weight measure
\begin{equation}
  \rho_0 \equiv \frac{1}{2}
  \left(
  \left|A_{\rm{loc}}(\omega=\delta)\right|
  +
  \left|A_{\rm{loc}}(\omega=-\delta)\right|
  \right), \label{eq:rho_0}
\end{equation}
where $\delta=0.002U$. As shown in Fig.\ \ref{fig:bg_spec}a, in this first regime of the Bose glass, the spectral weight $\rho_0$ for finite hopping is very close to the atomic limit result. In fact, the spectral function (Fig.\ \ref{fig:bg_spec}d) only differs from the atomic limit result at the edges of the upper Hubbard band, corresponding to the excitation $(1\rightarrow 2)$, indicating a strong localization around empty sites with large values of $\eta_i$.

The situation changes abruptly for $D \gtrsim U-\mu$ ($D/J \gtrsim 84$). As doubly occupied sites are activated by the disorder (see Fig.\ \ref{fig:bg_swe}d), the density increases  (Fig.\ \ref{fig:bg_swe}a), and so does the kinetic energy (Fig.\ \ref{fig:bg_swe}b), indicating an increase of non-local kinetic processes. The additional doublons lead to a substantial increase in interaction energy (Fig.\ \ref{fig:bg_swe}c), which dominates over the kinetic energy. One would therefore naively expect a better agreement between the spectral functions computed with SFA3 and in the atomic limit.
This is however not the case for the spectral weight around zero frequency $\rho_0$ which increases abruptly at $D/J \approx 84$, see Fig.\ \ref{fig:bg_spec}a, deviating markedly from the atomic limit prediction.
The appearance of doubly-occupied sites in the atomic limit drives additional excitations $(2\rightarrow 1)$ and $(2\rightarrow 3)$ in the spectral function (Fig.\ \ref{fig:bg_spec}e), which overlap with other excitations, leading to additional ``bands'' composed of multiple resonating excitations, see e.g.\ $(1\rightarrow 0,2\rightarrow 1)$ at low negative frequencies in Fig.\ \ref{fig:bg_spec}e.  

It is at the edges of these new ``bands'' that the spectral function is strongly peaked showing a considerable difference with respect to the atomic-limit spectral function, indicating delocalization of quasi-particles and quasi-holes in the vicinity of the rare sites with occupation $n>1$ (i.e.\ occupation $2$ in the atomic limit).
However, in the Bose glass discussed here, there is no global superfluid response, as the sites contributing to these peaks are rare.
Instead the physics is described by the notion of isolated superfluid lakes \cite{Gur_09} around rare sites with particularly low local potential. 

In this regime (denoted by the grey area in Fig.\ \ref{fig:bg_swe}), the non-local Green's function of SFA3 develops a simple pole at zero Matsubara frequency, which can be integrated out when computing local quantities such as the local Green's function, see \ref{app:pole} for details. Whence, the self-energy functional and local observables can still be evaluated in this regime. In a homogeneous system, such a pole would indicate an instability towards spontaneous $U(1)$-symmetry-breaking and the particles would condense. Here, however, it is only the rare sites with $n>1$ that contribute to the pole, not allowing for a global condensate. The pole therefore implies the presence of isolated quasi-condensates on the lattice, which can have different $U(1)$ phases and therefore do not allow for global phase-coherence (i.e.\ a finite superfluid response).
These highly non-local processes in the vicinity of a superfluid phase transition cannot be expected to be fully captured by the self-energies of a local reference system with translationally invariant variational parameters, leading to a deviation in the SFA3 kinetic energy with respect to the numerically exact QMC data in Fig.\ \ref{fig:bg_swe}b. This was also the case in close proximity to phase transitions in SFT  \cite{BSFT} and BDMFT  \cite{BDMFT,BDMFT1} calculations in the clean BHm.

For even stronger disorder, the situation changes
when the number of doubly occupied sites in the atomic limit (proportional to $r_2$ in Fig.\ \ref{fig:bg_swe}d) increases further: the background containing more and more strongly interacting doublons (see the increase of interaction energy in Fig.\ \ref{fig:bg_swe}c) makes it harder for particles to delocalize. This can be observed in the kinetic energy of Fig.\ \ref{fig:bg_swe}b, which decreases again as the particles localize. The same behavior can also be seen in the spectral function of Fig.\ \ref{fig:bg_spec}f, where the bands involving highly occupied sites increase in width, but are much closer to the atomic limit results. The zero frequency spectral weight $\rho_0$ decreases accordingly, as shown in Fig.\ \ref{fig:bg_spec}a.

When the disorder is strong enough to activate triplon occupancies, the behaviour changes once more.
The kinetic energy (Fig.\ \ref{fig:bg_swe}b) increases as the particles delocalize around the rare triply occupied sites, and so does the interaction energy (Fig.\ \ref{fig:bg_swe}c). The number of doubly occupied sites on the other hand decreases and $r_2=r_1$ (see Fig.\ \ref{fig:bg_swe}d). This behavior arises naturally from the probabilities $r_n$ of Sec.\ \ref{Local_Occ}:
once $D>\eta_{\rm{max}}(n)$ and  $D>-\eta_{\rm{min}}(n)$, the probability of finding a site with local occupation $n>0$ is given by the particle-number-independent value $r_n=U/2D$ (as in this case $r_1$ and $r_2$).

In summary, our results show that at fixed interaction $U/J$ (and chemical potential $\mu/U$) the strongly interacting Bose glass as a function of $D$ is described by the subsequent activation of local occupations $n$. As these occupations accumulate, the interaction energy increases, driving the phase towards the atomic limit. This is however not the case whenever a particular higher occupation number is very rare (i.e. if $0<r_n\ll 1$ for some local occupation $n>0$): in this case the particles tend to delocalize and form superfluid lakes \cite{Gur_09} around these rare highly occupied sites.

% --------------------------------------------------------------------
\subsection{Strongly-interacting phase transition}
\label{sec:SF_Finger}
% --------------------------------------------------------------------
%
\begin{figure}
\centering
\includegraphics[width=300pt]{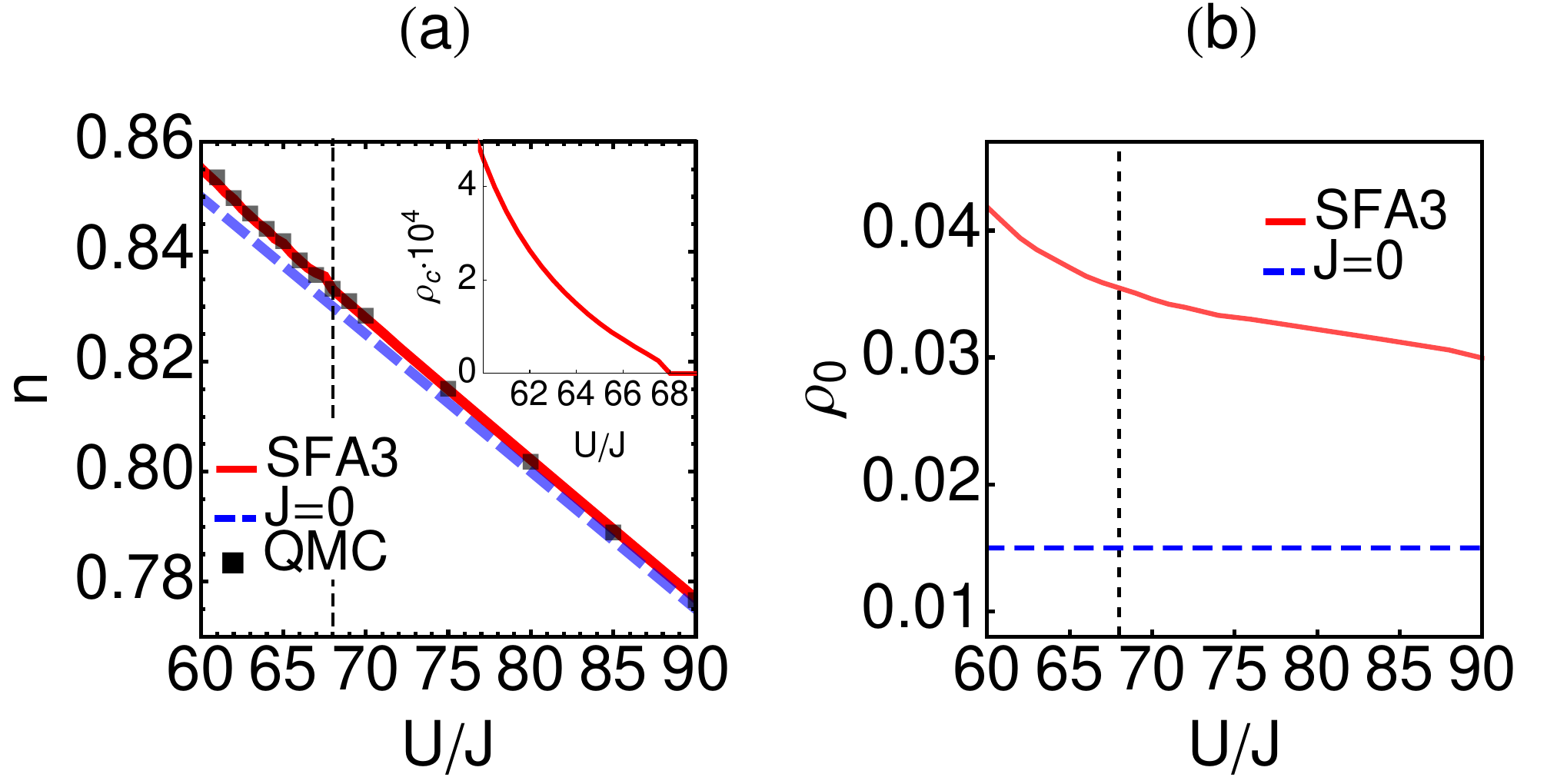}
  \caption{\label{fig:pt_swe} Strongly-interacting superfluid to Bose glass phase transition at $D/J=100$, $\mu/U=0.25$ and $T/J=0.1$. (a) Density $n$ and condensate-density $\rho_c$ (inset, rescaled by $10^4$) as a function of $U/J$ computed with SFA3 (red), QMC (black squares) and in the atomic limit (blue dashed). (b) Spectral weight around $\omega=0$, $\rho_0$ [see Eq.\ (\ref{eq:rho_0})], as a function of $U/J$ computed with SFA3 (red) and in the atomic limit (blue dashed). The vertical dashed line indicates the phase transition between the superfluid and the Bose glass phase in SFA3.}
\end{figure}

The regime where the Bose glass exhibits superfluid lakes around doubly occupied sites surrounds the superfluid finger at large interactions in the phase diagram of Fig.\ \ref{fig:pha_da}. In fact, the lower edge of the superfluid finger strongly correlates with the line where doubly-occupied sites are activated in the atomic limit (see blue dashed line in Fig.\ \ref{fig:pha_da}). The strongly-interacting phase transition at fixed chemical potential is illustrated in Fig.\ \ref{fig:pt_swe}, where we show the superfluid to Bose glass phase transition as a function of $U/J$ at fixed chemical potential $\mu/U=0.25$ and disorder strength $D/J=100$. 

With decreasing interaction $U/J$, these superfluid lakes percolate and resonances between the low-energy excitations $(0\rightarrow 1)$ and $(1 \rightarrow 2)$ on neighboring sites (and between the corresponding particle-removal processes, see e.g.\ Fig.\ \ref{fig:bg_spec}e) favour the spontaneous breaking of $U(1)$-symmetry through a homogeneous condensate. The particles therefore eventually condense, driving the transition to the superfluid phase. As the sites contributing to the resonating low-energy excitations remain relatively rare, the condensate fraction and the correction of the density with respect to the atomic limit close to the phase transition are extremely low (with a condensate density on the order of $10^{-4}$, see inset of Fig.\ \ref{fig:pt_swe}a).

At density $n\approx 1$ and larger disorder strength, the increase of highly occupied sites is compensated by a proliferation of empty sites (see e.g.\ Fig.\ \ref{fig:bg_swe}d).
Thus, the probability of having neighboring sites with resonating low-energy excitations decreases, making the spontaneous breaking of $U(1)$-symmetry less likely.
The increased interaction energy and particle number fluctuations therefore suppress the superfluid phase, explaining the reentrant behavior of the superfluid finger at larger disorder (see Fig.\ \ref{fig:pha_da}).

% --------------------------------------------------------------------
\subsection{Superfluid phase}
\label{sec:SF_phase}
% --------------------------------------------------------------------
%
\begin{figure}
\centering
\includegraphics[width=300pt]{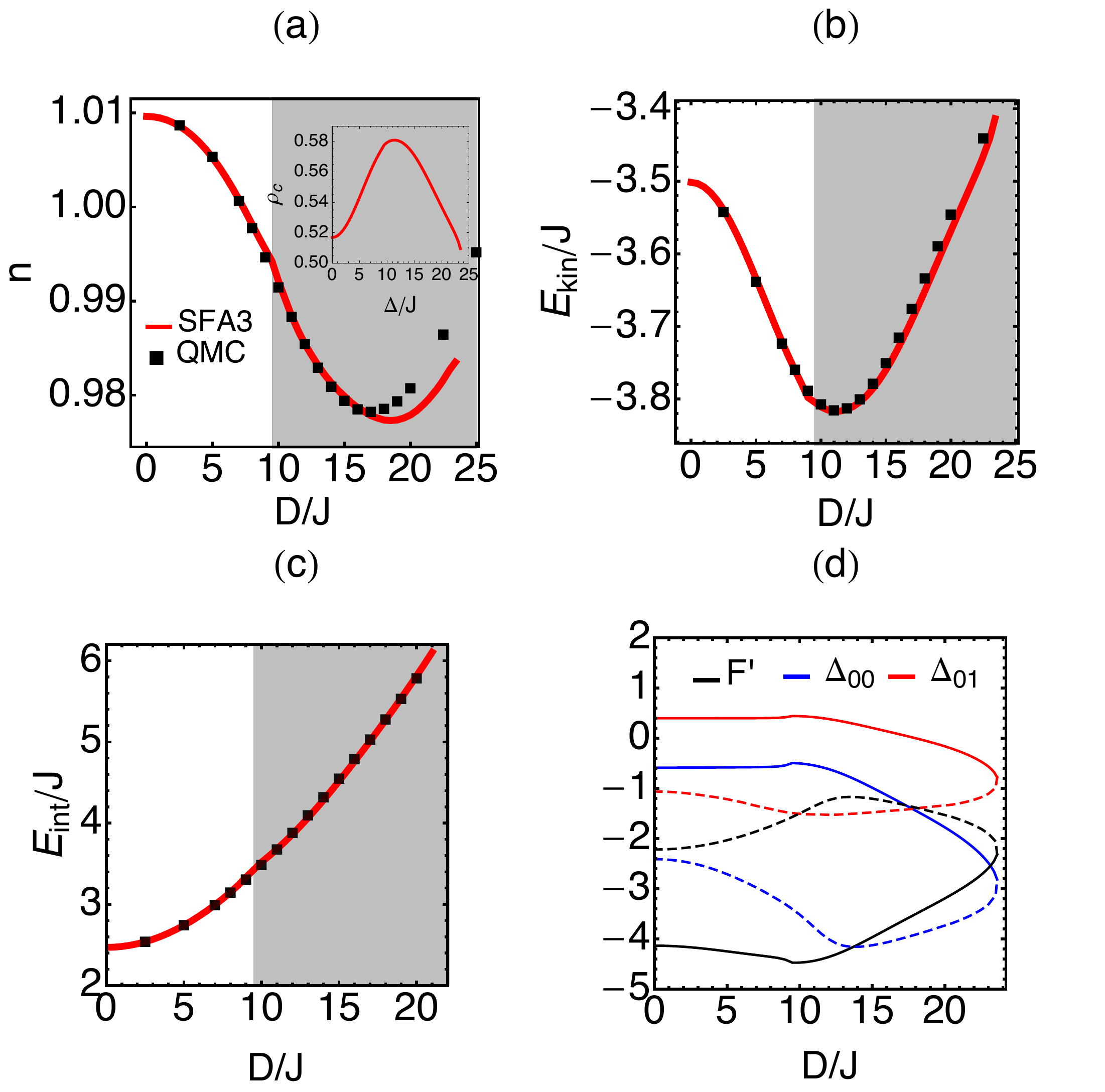}
  \caption{\label{fig:sf_swe} (a-c) Observables of the disordered BHm in the superfluid phase for $U/J=20$, $\mu/U=0.35$ and $T/J=0.1$, computed with SFA3 (red) and QMC (black squares). (a) Density $n$ and condensate density $\rho_c$ (inset) as a function of disorder strength $D/J$. (b) Kinetic energy per site $E_{\rm{kin}}/J$ as a function of disorder strength $D/J$. (c) Interaction energy per site $E_{\rm{int}}/J$ as a function of disorder strength $D/J$. The grey area indicates the region where the non-local Green's function of SFA3 develops a pole. (d) Variational parameters of the SFA3 calculation, $F'$ (black), $\Delta_{00}$ (blue) and $\Delta_{01}$ (red), as a function of disorder strength $D/J$. The solid lines indicate the stationary solution corresponding to panels (a-c), while the dashed line shows a metastable solution.}
\end{figure}

We now turn to lower interactions, i.e., deeper into the superfluid phase away from the superfluid finger. If $U/J$ is lower than the critical value of the clean system, the condensate density is much larger than in the superfluid finger, and the uncondensed background is no longer well described by the atomic limit.

In Fig.\ \ref{fig:sf_swe} we show a sweep of the thermodynamical observables as a function of $D/J$ deep in the superfluid phase at $U/J=20$ and $\mu/U=0.35$. At low disorder SFA3 shows excellent agreement with QMC, as the condensate density increases and the density decreases as a function of disorder (Fig.\ \ref{fig:sf_swe}a). As a consequence of the larger condensate fraction, the magnitude of the kinetic energy increases as well (Fig.\ \ref{fig:sf_swe}b). The interaction energy increases throughout the entire parameter range $0 \le D / J \le 25$, indicating increasing spatial particle number fluctuations (Fig.\ \ref{fig:sf_swe}c).

When the disorder becomes comparable to the single-particle bandwidth $ W= 2zJ$, these fluctuations reverse the trend of the condensate density which starts to decrease as a function of $D/J$.
It is at this point that also the kinetic energy starts to decrease and the non-local connected Green's function develops a pole at zero Matsubara frequency (see \ref{app:pole}).  As in the Bose glass (see Sec.\ \ref{sec:strong_BG}), the pole indicates the appearance of additional isolated quasi-condensates in the system: this is most likely related to the disorder inducing rare regions, explaining the decrease in condensate density and kinetic energy, and leading to a glassy behavior in the superfluid. 

Eventually, deeper in the glassy regime of the superfluid where the disorder dominates over both the interaction and the single-particle bandwidth, our SFA3 approach of having translationally-invariant variational parameters on the reference system becomes too simple to fully capture the groundstate behavior. In fact, the SFA3 results start to deviate from the QMC results, see Fig.\ \ref{fig:sf_swe}a. As shown in Fig.\ \ref{fig:sf_swe}d, eventually at $D/J\approx 2W$ the variational parameters of the stationary SFA3 solution join with a metastable solution with higher free energy through a saddle-node bifurcation \cite{Crawford:1991qy}, vanishing for larger disorder. 

A possibility to get around this problem, may be the introduction of a spatially modulated symmetry-breaking field on the reference system, as was also done in stochastic mean-field theory \cite{Stoch_MF0,Stoch_MF}. In SFT, this would however involve the inversion of a non-translationally-invariant connected Green's function, limiting us to very small system sizes, while we here want to analyze the thermodynamic limit.

% --------------------------------------------------------------------
\section{Conclusion}
\label{sec:conc}
% --------------------------------------------------------------------

In this work, we generalized the bosonic self-energy functional theory (SFT) to include quenched disorder. The derived formalism is a general framework for constructing non-perturbative approximations of disordered interacting bosonic lattice systems incorporating spontaneous $U(1)$-symmetry-breaking. We showed that the resulting SFT functional depends only on the self-energies of the disorder-averaged interacting one- and two-point propagator, the condensate and connected Green's function, respectively. The lattice self-energies can then be parametrized by the self-energies of a simpler exactly solvable reference system having the same interaction and disorder distribution as the original system. The resulting formalism is a general non-perturbative approach that contains disorder-averaged bosonic dynamical mean-field theory as a certain limit.
%
%, but can also be used as a more general variational scheme. 

We applied SFT in combination with a simple SFA3 reference system, consisting of a single bosonic mode with only three variational parameters, to the Bose-Hubbard model with local box disorder. The SFA3 results were compared to numerically exact path integral quantum Monte Carlo (QMC) results and analytic calculations in the atomic limit.

Our results in the strongly-interacting regime close to unit filling, showing excellent agreement with QMC, indicate that the Bose glass phase is characterized by different regimes as a function of the disorder strength $D$. With increasing $D$, sites with local occupations $n\neq 1$ appear as predicted by the atomic limit, leading to crossovers between different regimes whenever a new local occupation $n$ is activated by the disorder. While QMC has to resort to analytic continuation in order to compute dynamical quantities, through SFT we were able to compute spectral functions within SFA3 directly.

By systematically analyzing the local spectral function we observed that the bosons delocalize into superfluid lakes around highly occupied sites whenever these are particularly rare. In particular, we found that the transition from the strongly-interacting Bose glass to the strongly-interacting superfluid phase (which extends to values of the interaction which are much larger than in the clean system) is driven by the percolation of superfluid lakes which form around doubly occupied sites, leading to a small condensate fraction over a strongly localized background. As $D$ is further increased and the density of doublons increases accordingly, the particles are localized by the strongly interacting background, explaining the reentrant behavior of the superfluid phase. 

Due to the extremely low condensate fraction in the strongly-interacting superfluid, even though the numerical error is small, the phase boundaries observed with SFA3 are shifted with respect to the QMC results. Deeper in the superfluid phase (i.e.\ at lower interactions), our SFT results show excellent agreement with the QMC data as long as the disorder is smaller or comparable to the non-interacting bandwidth. In the strongly-disordered weakly-interacting regime, however, the restricted variational subspace of the SFA3 reference system employed in this work is no-longer capable to find a stationary solution.

As opposed to QMC, SFT does not suffer from a general sign problem in the presence of e.g.\ gauge fields \cite{Struck:2012aa, Greschner:2014aa, Goldman:2014aa}, or other complex Hamiltonian terms such as spin-orbit coupling \cite{Lin:2011aa, Struck:2014aa, Jimenez-Garcia:2015aa}. The formalism derived in this work therefore represents a promising tool for future studies of such complex systems in the presence of disorder. In particular, an extension to real-time dynamics, as has been done for disorder-free fermionic systems~\cite{Hofmann:2013aa,Hofmann:2016ab}, seems a promising route to study the elusive physics of many-body-localized systems and is left for future work.

\section*{Acknowledgments}
The authors would like to thank E. Altman, D. A. Huse, M. Knap, and T. Pfeffer for fruitful discussions and valuable input. DH and LP are supported  by  FP7/ERC  Starting  Grant  No.   306897  and FP7/Marie-Curie Grant No.  321918, HS is supported by the Swiss National Science Foundation through NCCR MARVEL.

\appendix

% --------------------------------------------------------------------
\section{Derivation of the disorder averaged self-energy functional}
\label{App:T_deriv}
% --------------------------------------------------------------------

%By insertion of the averaged Dyson equations [Eq.\ (\ref{eq:Sigm_P})] in the definitions of the averaged propagators $\bar{\bPhi}$ and $\bar{\bG}$ we obtain the relations
%
%\begin{equation}
%  \bar{\bPhi} \equiv
%  \langle \hat{\bPhi}_{\heta} \rangle_P
%  =
%  \Big\langle
%      [ \bar{\bG}^{-1} + \bar{\bS} - \heta ]^{-1}
%      ( \bar{\mbf{S}} - \mbf{S}_{\heta} + [\bar{\bG}^{-1} + \bar{\bS}] \bar{\bPhi} )
%      \Big\rangle_P
%  \, , \label{eq:AvgPhiFunc}
%\end{equation}
%
%\begin{eqnarray}
%  \bar{\bG} & \equiv &
%  \langle \hat{\bG}_{\heta} - \hat{\bPhi}_{\heta} \hat{\bPhi}_{\heta}^\dagger \rangle_P
%  + \bar{\bPhi} \bar{\bPhi}^\dagger
%   =
%  \Big\langle
%      [ \bar{\bG}^{-1} + \bar{\bS} - \heta - \bS_{\heta} ]^{-1}
%  \\ & & -
%      [ (\bar{\bG}^{-1} + \bar{\bS} - \heta )^{-1}
%        ( \bar{\mbf{S}} - \mbf{S}_{\heta} + [\bar{\bG}^{-1} + \bar{\bS}] \bar{\bPhi} )
%      ]^2
%  \Big\rangle_P
%  . \label{eq:AvgGFunc}
%\end{eqnarray}
%
%The concomitant solution of Eq.\ (\ref{eq:AvgPhiFunc}) and (\ref{eq:AvgGFunc}) implicitly defines the two universal functionals $\hat{\bar{\bPhi}}[ \bar{\mbf{S}}, \bar{\bS}, \{ \mbf{S}_{\heta}, \bS_{\heta} \}] = \bar{\bPhi}$ and $\hat{\bar{\bG}}[ \bar{\mbf{S}}, \bar{\bS}, \{ \mbf{S}_{\heta}, \bS_{\heta} \}] = \bar{\bG}$.

In order to construct approximations using the disorder averaged propagators we now seek a functional that is equal to the disorder average of the self energy functional $\Big\langle \SE [ \mbf{S}_{\heta}, \bS_{\heta} ] \Big\rangle_P$ but that is defined in the extended space of both averaged and explicit self-energies $\bar{\mbf{S}}, \bar{\bS}, \{ \mbf{S}_{\heta}, \bS_{\heta} \}$ and stationary at the physical solution in all self energies.

An ansatz that fulfills equality at the physical self-energies is Eq.\ (\ref{eq:DisorderAveragedExtendedSFT}).
%
%\begin{eqnarray}
%  \ASE[\bar{\mbf{S}}, \bar{\bS}, \{ \mbf{S}_{\heta}, \bS_{\heta} \} ]
%  \equiv &
%  \frac{1}{2} (\bF - \bar{\mbf{S}})^\dagger \bG_{\mbf{t} 0 0} (\bF - \bar{\mbf{S}})
%  + \frac{1}{2} \Tr \ln [ - (\bG^{-1}_{\mbf{t} 0 0} - \bar{\bS})] \nonumber
%  \\
%  &+
%  %\hat{\mathcal{T}}_{P}
%  \ALWSE
%      [\bar{\mbf{S}}, \bar{\bS}, \{ \mbf{S}_{\heta}, \bS_{\heta} \} ]
%  +
%  \left\langle
%  \LWSE[\mbf{S}_{\heta}, \bS_{\heta}]
%  \right\rangle_P
%  \, ,
%  \label{eq:ASE_Phys_Value}
%\end{eqnarray}
%
%where the universal functional $\ALWSE$ is given by
%
%\begin{eqnarray}
%   \ALWSE [\bar{\mbf{S}}, \bar{\bS}, \{ \mbf{S}_{\heta}, \bS_{\heta} \} ]
%  \equiv
%
%  - \frac{1}{2}
%  \hat{\bar{\bPhi}}^\dagger
%  ( \hat{\bar{\bG}}^{-1} + \bar{\bS} )
%  \hat{\bar{\bPhi}}
%  -
%  \frac{1}{2}
%  \left\langle
%    \hat{\bPhi}_{\heta}^{\dagger}
%    ( \hat{\bar{\bG}}^{-1} + \bar{\bS} - \heta )
%    \hat{\bPhi}_{\heta}^{}
%  \right\rangle_P
%  \nonumber
%  \\
%   +
%  \left\langle
%  ( \bar{\mbf{S}} - \mbf{S}_{\heta} + \bar{\bPhi} [ \hat{\bar{\bG}}^{-1} + \bar{\bS} ] )
%  \hat{\bPhi}_{\heta}
%  \right\rangle_P
%   +
%  \frac{1}{2}
%  \left\langle
%  \Tr \ln [ - ( \hat{\bar{\bG}}^{-1} + \bar{\bS} - \heta - \bS_{\heta} ) ]
%  \right\rangle_P
%  \nonumber
%  \\ -
%  \frac{1}{2} \Tr \ln ( - \hat{\bar{\bG}}^{-1} )
%  \,  \label{eq:ALWSE_Def}.  
%\end{eqnarray}
%
Repeated application of the Dyson equations [Eqs.\ (\ref{eq:Sigm_P}) and (\ref{eq:D1})] in Eq.\ (\ref{eq:ALWSE_Def}) at the physical self-energies gives
\begin{eqnarray}
   \ALWSE [\bar{\mbf{S}}_{\bF\bt PV}, \bar{\bS}_{\bF\bt PV}, \{ \mbf{S}_{\bF\bt\heta V}, \bS_{\bF\bt\heta V} \} ]
  =- \frac{1}{2} (\bF - \bar{\mbf{S}}_{\bF\bt PV})^\dagger \bG_{\mbf{t} 0 0} (\bF - \bar{\mbf{S}}_{\bF\bt PV})\nonumber \\
  - \frac{1}{2} \Tr \ln [ - (\bG^{-1}_{\mbf{t} 0 0} - \bar{\bS}_{\bF\bt PV})] 
  +\frac{1}{2}\left\langle (\bF - \bar{\mbf{S}}_{\bF\bt\heta V})^\dagger \bG_{\mbf{t} \heta 0} (\bF - \bar{\mbf{S}}_{\bF\bt\heta V})\right\rangle_P
 \nonumber\\
   + \frac{1}{2}\left\langle \Tr \ln [ - (\bG^{-1}_{\mbf{t} \heta 0} - \bar{\bS}_{\bF\bt\heta V})]\right\rangle_P
  \, , \label{eq:stat_Tpv}
\end{eqnarray}
and therefore
\begin{equation}
  \ASE[\bar{\mbf{S}}_{\bF\bt PV}, \bar{\bS}_{\bF\bt PV}, \{ \mbf{S}_{\bF\bt\heta V}, \bS_{\bF\bt\heta V} \} ]
  =
  \Big\langle \SE [ \mbf{S}_{\bF\bt\heta V}, \bS_{\bF\bt\heta V} ] \Big\rangle_P
  \, ,
\end{equation}
whence the disorder averaged self-energy functional gives the physical disorder averaged free energy at stationarity.
To show stationarity of $\ASE$ we consider the variations of the universal functional $\ALWSE$
\begin{eqnarray}
  \delta_{\bar{\mbf{S}}^\dagger} \ALWSE
  = &
  \langle \hat{\bPhi}_{\heta} \rangle_P
  %\nonumber \\
  %& +
  +
  ( \delta_{\bar{\mbf{S}}^\dagger} \hat{\bar{\bPhi}}^\dagger )
  [ \hat{\bar{\bG}}^{-1} + \bar{\bS} ]
  \left[
    \langle \hat{\bPhi}_{\heta} \rangle_P - \hat{\bar{\bPhi}}
    \right]
   \\ \nonumber
  & +
  ( \delta_{\bar{\mbf{S}}^\dagger} \hat{\bar{\bG}}^{-1} )
  \left[
    - \frac{1}{2} \hat{\bar{\bPhi}} \hat{\bar{\bPhi}}^\dagger
    - \frac{1}{2} \langle \hat{\bPhi}_{\heta} \hat{\bPhi}_{\heta}^\dagger \rangle_P
    + \langle \hat{\bPhi}_{\heta} \rangle_P \hat{\bar{\bPhi}}^\dagger 
    + \frac{1}{2} \langle \hat{\bG}_{\heta} \rangle_P
    - \frac{1}{2} \hat{\bar{\bG}}
    \right]
\end{eqnarray}

\clearpage

\begin{eqnarray}
  \delta_{\bar{\bS}} \ALWSE
  = &
  - \frac{1}{2} \hat{\bar{\bPhi}} \hat{\bar{\bPhi}}^\dagger
  - \frac{1}{2} \langle \hat{\bPhi}_{\heta}\hat{\bPhi}_{\heta}^\dagger \rangle_P
  +
  \langle \hat{\bPhi}_{\heta} \rangle_P
  \hat{\bar{\bPhi}}^\dagger
  + \frac{1}{2} \langle \hat{\bG}_{\heta} \rangle_P
  \nonumber \\
  & +
  ( \delta_{\bar{\bS}} \hat{\bar{\bPhi}}^\dagger )
  [ \hat{\bar{\bG}}^{-1} + \bar{\bS} ]
  \left[
    \langle \hat{\bPhi}_{\heta} \rangle_P - \hat{\bar{\bPhi}}
    \right]
   \\ \nonumber
  & +
  ( \delta_{\bar{\bS}} \hat{\bar{\bG}}^{-1} )
  \left[
    - \frac{1}{2} \hat{\bar{\bPhi}} \hat{\bar{\bPhi}}^\dagger
    - \frac{1}{2} \langle \hat{\bPhi}_{\heta} \hat{\bPhi}_{\heta}^\dagger \rangle_P
    + \langle \hat{\bPhi}_{\heta} \rangle_P \hat{\bar{\bPhi}}^\dagger 
    + \frac{1}{2} \langle \hat{\bG}_{\heta} \rangle_P
    - \frac{1}{2} \hat{\bar{\bG}}
    \right]  
\end{eqnarray}
\begin{eqnarray}
  \delta_{\mbf{S}_{\heta}^\dagger} \ALWSE
  = &
  - P(\heta) \hat{\bPhi}_{\heta}
 +
  ( \delta_{\mbf{S}_{\heta}^\dagger} \hat{\bar{\bPhi}}^\dagger )
  [ \hat{\bar{\bG}}^{-1} + \bar{\bS} ]
  \left[
    \langle \hat{\bPhi}_{\heta} \rangle_P - \hat{\bar{\bPhi}}
    \right]
   \\ \nonumber
  & +
  ( \delta_{\mbf{S}_{\heta}^\dagger} \hat{\bar{\bG}}^{-1} )
  \left[
    - \frac{1}{2} \hat{\bar{\bPhi}} \hat{\bar{\bPhi}}^\dagger
    - \frac{1}{2} \langle \hat{\bPhi}_{\heta} \hat{\bPhi}_{\heta}^\dagger \rangle_P
    + \langle \hat{\bPhi}_{\heta} \rangle_P \hat{\bar{\bPhi}}^\dagger 
    + \frac{1}{2} \langle \hat{\bG}_{\heta} \rangle_P
    - \frac{1}{2} \hat{\bar{\bG}}
    \right]
  \\ \nonumber
  & +
  ( \delta_{\mbf{S}_{\heta}^\dagger} \hat{\bPhi}_{\heta}^\dagger ) P(\heta)
  \left[
    \bar{\mbf{S}}
    + [ \hat{\bar{\bG}}^{-1} + \bar{\bS} ] \hat{\bar{\bPhi}}
    - \mbf{S}_{\heta}
    - [\hat{\bar{\bG}}^{-1} + \bar{\bS} - \heta] \hat{\bPhi}_{\heta}
    \right]
\end{eqnarray}
\begin{eqnarray}
  \delta_{\bS_{\heta}} \ALWSE
  = &
  - \frac{1}{2} P(\heta) \hat{\bG}_{\heta}
   +
  ( \delta_{\bS_{\heta}} \hat{\bar{\bPhi}}^\dagger )
  [ \hat{\bar{\bG}}^{-1} + \bar{\bS} ]
  \left[
    \langle \hat{\bPhi}_{\heta} \rangle_P - \hat{\bar{\bPhi}}
    \right]
   \\ \nonumber
  & +
  ( \delta_{\bS_{\heta}} \hat{\bar{\bG}}^{-1} )
  \left[
    - \frac{1}{2} \hat{\bar{\bPhi}} \hat{\bar{\bPhi}}^\dagger
    - \frac{1}{2} \langle \hat{\bPhi}_{\heta} \hat{\bPhi}_{\heta}^\dagger \rangle_P
    + \langle \hat{\bPhi}_{\heta} \rangle_P \hat{\bar{\bPhi}}^\dagger 
    + \frac{1}{2} \langle \hat{\bG}_{\heta} \rangle_P
    - \frac{1}{2} \hat{\bar{\bG}}
    \right]
  \\ \nonumber
  & +
  ( \delta_{\bS_{\heta}} \hat{\bPhi}_{\heta}^\dagger ) P(\heta)
  \left[
    \bar{\mbf{S}}
    + [ \hat{\bar{\bG}}^{-1} + \bar{\bS} ] \hat{\bar{\bPhi}}
    - \mbf{S}_{\heta}
    - [\hat{\bar{\bG}}^{-1} + \bar{\bS} - \heta] \hat{\bPhi}_{\heta}
    \right]
\end{eqnarray}
Using the definitions of the averaged propagators [Eq.\ (\ref{eq:phi_funct_P})] this reduces to
\begin{eqnarray}
  \delta_{\bar{\mbf{S}}^\dagger} \ALWSE
  =
  \hat{\bar{\bPhi}}
  \, , \quad
  \delta_{\mbf{S}_{\heta}^\dagger} \ALWSE
  = 
  - P(\heta) \hat{\bPhi}_{\heta}
  +
  ( \delta_{\mbf{S}_{\heta}^\dagger} \hat{\bPhi}_{\heta}^\dagger ) P(\heta)
  \mathcal{Q}_{\heta}
  \label{eq:S_var_ALWSE}
  \\
  \delta_{\bar{\bS}} \ALWSE
  =
  \hat{\bar{\bG}}
  \, , \quad
  \delta_{\bS_{\heta}} \ALWSE
  = 
  - \frac{1}{2} P(\heta) \hat{\bG}_{\heta}
  +
  ( \delta_{\bS_{\heta}} \hat{\bPhi}_{\heta}^\dagger ) P(\heta)
  \mathcal{Q}_{\heta}
  \, ,
  \label{eq:Sigma_var_ALWSE}
\end{eqnarray}
where
\begin{equation}
  \mathcal{Q}_{\heta} \equiv
  \bar{\mbf{S}}
  + [ \hat{\bar{\bG}}^{-1} + \bar{\bS} ] \hat{\bar{\bPhi}}
  - \mbf{S}_{\heta}
  - [\hat{\bar{\bG}}^{-1} + \bar{\bS} - \heta] \hat{\bPhi}_{\heta}
  \, .
\end{equation}
The term $\mathcal{Q}_{\heta}$ corresponds to $\mathcal{Q}_{\heta} = \bF - \bF=0$ when the one point Dyson equations [Eqs. (\ref{eq:Sigm_P}) and (\ref{eq:D1})] are fulfilled. Hence, the $\delta \hat{\bPhi}_{\heta}^\dagger$ variations in Eqs. (\ref{eq:S_var_ALWSE}) and (\ref{eq:Sigma_var_ALWSE}) vanish at stationarity (i.e.\ at the physical self-energies).

% --------------------------------------------------------------------
\section{Canceling functional derivatives of $\ALWSE$}
\label{App:Univ}
% --------------------------------------------------------------------

At stationarity the expression of the $\ALWSE$ functional in Eq.\ (\ref{eq:stat_Tpv}) contains expressions in terms of $\bG_{\mbf{t} \heta 0}$ and $\bF$, so one might wonder if there is no implicit dependence on the free propagators of the system. In order to check that this is not the case, we rewrite Eq.\ (\ref{eq:stat_Tpv}) in terms of $\bG_{\mbf{t} 0 0}$ and $\bF$ as
\begin{eqnarray}
 \ALWSE [\bar{\mbf{S}}_{\bF\bt PV}, \bar{\bS}_{\bF\bt PV}, \{ \mbf{S}_{\bF\bt\heta V}, \bS_{\bF\bt\heta V} \}] = 
 \frac{1}{2}\left\langle\Tr\ln\left[-\left(\bG^{-1}_{\mbf{t} 0 0}-\heta-\bS_{\bF\bt\heta V}\right)\right]\right\ranp \nonumber \\
  -\frac{1}{2}\Tr\ln\left[-\left(\bG^{-1}_{\mbf{t} 0 0}-\bar{\bS}_{\bF\bt PV}\right)\right] -\frac{1}{2} (\bF - \bar{\mbf{S}}_{\bF\bt PV})^\dagger \bG_{\mbf{t} 0 0} (\bF - \bar{\mbf{S}}_{\bF\bt PV}) \nonumber\\ 
+ \frac{1}{2} \left\langle(\bF - \mbf{S}_{\bF\bt\heta V})^\dagger \left[\bG^{-1}_{\mbf{t} 0 0}-\heta\right]^{-1} (\bF - \mbf{S}_{\bF\bt\heta V})\right\ranp .
\end{eqnarray}

The variation in $\bG^{-1}_{\mbf{t} 0 0}$ yields
\begin{eqnarray}
\frac{\delta \ALWSE [\bar{\mbf{S}}_{\bF\bt PV}, \bar{\bS}_{\bF\bt PV}, \{ \mbf{S}_{\bF\bt\heta V}, \bS_{\bF\bt\heta V} \}] }{\delta\bG^{-1}_{\mbf{t} 0 0}} =\frac{1}{2}\left\langle\Tr\left[\bG^{-1}_{\mbf{t} {\heta} 0}-\bS_{\bF\bt\heta V}\right]^{-1}\right\ranp \nonumber \\
 -\frac{1}{2}\Tr\left[\bG^{-1}_{\mbf{t} 0 0}- \bar{\bS}_{\bF\bt PV}\right]^{-1}-\frac{1}{2} \left\langle(\bF - \mbf{S}_{\bF\bt\heta V})^\dagger \left[\bG^{}_{\mbf{t} {\heta} 0}\right]^{2} (\bF - \mbf{S}_{\bF\bt\heta V})\right\ranp\nonumber \\
  +\frac{1}{2} (\bF - \bar{\mbf{S}}_{\bF\bt PV})^\dagger \left[\bG^{}_{\mbf{t} 0 0}\right]^{2} (\bF - \bar{\mbf{S}}_{\bF\bt PV}),
\end{eqnarray}
which by the short-hand notations introduced in Eqs.\ (\ref{eq:phi_funct_P}),  and (\ref{eq:G_funct_P}) can be rewritten as
\begin{eqnarray}
\frac{\delta \ALWSE [\bar{\mbf{S}}_{\bF\bt PV}, \bar{\bS}_{\bF\bt PV}, \{ \mbf{S}_{\bF\bt\heta V}, \bS_{\bF\bt\heta V} \} ] }{\delta\bG^{-1}_{\mbf{t} 0 0}} =& \frac{1}{2}\Tr\left\langle{\bG}_{\heta}\right\ranp  -\frac{1}{2} \left\langle{\bPhi}_{\heta}^\dagger {\bPhi}_{\heta}\right\ranp\nonumber \\ &- \frac{1}{2}\Tr  {\bar{\bG}}
+\frac{1}{2} {\bar{\bPhi}}^\dagger{\bar{\bPhi}} =0,
\label{der_2}
\end{eqnarray}
where we have used that the trace and the arithmetic average commute, i.e. $\frac{1}{2}\left\langle\Tr{\bG}_{\heta}\right\ranp=\frac{1}{2}\Tr\left\langle{\bG}_{\heta}\right\ranp$.
Note that -- as opposed to the arithmetical average -- the geometrical average used in the context of fermionic DMFT \cite{Dobro_PRL_97,Dobro_PRL_03,Byc_PRL_05,Byc_PRB_05} would not commute with the trace operator $\Tr$ in Eq.\ (\ref{der_2}) and therefore break the universality of the functional $\ALWSE$. As pointed out in Ref.\ \cite{Pot_Dis} for fermions, the geometrical average introduced in DMFT, therefore appears to be incompatible with SFT.

The variation of $\ALWSE$ in $\bF$ yields
\begin{equation}
\frac{\delta\ALWSE [\bar{\mbf{S}}_{\bF\bt PV}, \bar{\bS}_{\bF\bt PV}, \{ \mbf{S}_{\bF\bt\heta V}, \bS_{\bF\bt\heta V} \} ] }{\delta \bF^{\dagger}} = \bG^{}_{\mbf{t} 0 0}(\bF - \bar{\mbf{S}}_{\bF\bt PV}) - \left \langle\bG^{}_{\mbf{t} {\heta} 0}(\bF -\mbf{S}_{\bF\bt\heta V})\right\ranp,\nonumber
%\label{der_2}
\end{equation}
which using Eq.\ (\ref{eq:phi_funct_P}) can be rewritten as
\begin{equation}
\frac{\delta\ALWSE [\bar{\mbf{S}}_{\bF\bt PV}, \bar{\bS}_{\bF\bt PV}, \{ \mbf{S}_{\bF\bt\heta V}, \bS_{\bF\bt\heta V} \} ] }{\delta \bF^{\dagger}} ={\bar{\bPhi}}- \left \langle {\bPhi}_{\heta}\right\ranp=0.\nonumber
\end{equation}
$\ALWSE$ is therefore completely independent of both $\bF$ and $\bG^{-1}_{\mbf{t} 0 0}$ also at stationarity.

% ----------------------------------------------------------------------
\section{Disorder-averaged bosonic dynamical mean-field theory limit}
\label{sec:Dis_BDMFT}
% ----------------------------------------------------------------------

Disordered-averaged SFT for bosons has disorder-averaged bosonic dynamical mean-field theory (BDMFT) as a certain limit. In its simplest form, disorder-averaged BDMFT is restricted to site-local disorder $\heta$ and site-local interaction $\hat{V}$.

%To show that the disorder averaged bosonic dynamical mean-field theory (BDMFT) is a certain limit of disordered self-energy functional theory we have to constrain the interaction $\hat{V}$ and disorder $\eta$ to be site-local
%
%\begin{equation}
%  \eta_{ij} = \delta_{ij} \eta_{i}
%  \, .
%\end{equation}

In this case, disorder-averaged BDMFT is obtained from SFT by restricting the reference systems
free propagator to be site-local, i.e.
\begin{equation}
\bDelta^{i\nu}_{j\nu'}(\tau, \tau') = \delta_{ij} [\bDelta_{i}]^{\nu}_{\nu'}(\tau, \tau')
  \, ,\nonumber
\end{equation}
where $i,j$ are the site-, and $\nu,\nu'$ the Nambu indices. The imaginary time retardation in $\bDelta(\tau, \tau')$, however, remains completely general.

%The reference system free propagator must also be constrained to be site local
%
%\begin{equation}
%  [\bG_{\Delta'\eta 0}^{-1}]_{ij} = \delta_{ij} [\bG_{\Delta'\eta 0}^{-1}]_{ii}
%\end{equation}
%%
%which constrains $\Delta'$ to be local, while the retardation still is completely general, by hybridizing each site with a set of non-interacting bath sites, see Ref.\ \cite{BSFT} for details.
%
The reference systems local bare propagator $\bG_{\bDelta \heta 0}$ and interaction give rise to a purely local self-energy
\begin{equation}
  \left[\bS_{\bF'\bDelta\heta V}\right]_{ij}  = \delta_{ij} \left[\bS_{\bF'\bDelta\heta V}\right]_{ii} \, ,\quad
  \left[\bar{\bS}_{\bF'\bDelta PV}\right]_{ij}  = \delta_{ij} \left[\bar{\bS}_{\bF'\bDelta P V}\right]_{ii}
  \, ,
\end{equation}
and the self-energy variations of $\SFT$ [Eqs.\ (\ref{eq:dSFTdS}) and (\ref{eq:dSFTdSigma})] reduce to the disorder-averaged BDMFT self-consistency conditions
\begin{equation}
  \bG_{\bDelta 00}(\bF' - \bar{\mbf{S}}_{\bF'\bDelta} )
  -
  \bG_{\mbf{t} 00}(\bF - \bar{\mbf{S}}_{\bF'\bDelta} )
  =
  \bG_{\mbf{t}00}
  \left[
    (\bG_{\mbf{t}00}^{-1} - \bG_{\bDelta00}^{-1}) \bar{\bPhi}_{\bF'\bDelta} + \bF' - \bF
    \right]
  = 0
  \, ,
\end{equation}
\begin{equation}
  [(\bG_{\bDelta00}^{-1})_{ii} - \bar{\bS}_{\bF'\bDelta}]^{-1}
  -
  [\bG_{\mbf{t}00}^{-1} - \bar{\bS}_{\bF'\bDelta}]_{ii}^{-1}
  = 0
  \, ,\nonumber
\end{equation}
which can be fulfilled exactly by the retarded $\bDelta(\tau, \tau')$, and can be simplified to
\begin{equation}
\bar{\bG}_{ii}  = \bar{\bG}_{\bF'\bDelta PV} \, , \quad
\bar{\bPhi}   = \bar{\bPhi}_{\bF'\bDelta PV}
  \, ,
\end{equation}
where $\bar{\bPhi} $ and $\bar{\bG}_{ii}$ are the disorder-averaged condensate and local connected Green's function of the lattice, while $\bar{\bPhi}_{\bF'\bDelta PV}$ and $\bar{\bG}_{\bF'\bDelta PV}$ are the disorder-averaged condensate and connected Green's function of the reference system. This is therefore the standard BDMFT self-consistency condition of clean systems \cite{BDMFT,BDMFT1}, where the propagators of the clean system have been replaced by their disorder-averaged counterparts, which for the case of uncorrelated disorder discussed in \ref{sec:uncorrelated} are translationally invariant.

% --------------------------------------------------------------------
\section{Uncorrelated disorder: translational invariance of the arithmetic average}
\label{sec:uncorrelated}
% --------------------------------------------------------------------

In the following we will specialize the formalism derived in Sec.\ \ref{sec:theory} by assuming that the disorder is distributed according to an uncorrelated and translationally invariant probability distribution, i.e.
\begin{equation}
  P(\heta)=\prod_{ij}p_{i-j}(\eta_{ij}) \label{eq:uncor_P}
  \, ,
\end{equation}
where the product goes over all site-indices $i,j$, and the distribution $p_{i-j}$ depends only on the relative distance between the sites $i$ and $j$. As we will see this enables us to simplify the reference system considerably due to the translational invariance of disorder-averaged observables.

The interacting propagators at a given disorder configuration $\heta$ can be computed directly by
\begin{equation}
\bG_{\bF' \bDelta \heta V}(\tau-\tau') = -\langle \mathcal{T}\bbb(\tau)\bbc(\tau')\rangle_{{\heta}}+\langle \mbf{b}\rangle_{\heta}\langle \mbf{b}^{\dagger} \rangle_{\heta} \, , \quad
\bP_{\bF' \bDelta \heta V} = \langle \mbf{b}\rangle_{\heta} \label{eq:Pprime_eta}
\, ,
\end{equation}
where $\mathcal{T}$ is the time-ordering operator and $\langle \dots\rangle_{\heta}$ means taking the expectation value with respect to the reference system with disorder configuration $\heta$.

Using Eq.\ (\ref{eq:Pprime_eta}) further enables the computation of the fixed-disorder self-energies through Eq.\ (\ref{eq:D1}).
The propagators $\bG_{\bF' \bDelta \heta V}$, $\bP_{\bF' \bDelta \heta V}$, and the corresponding self-energies $\bS_{\bF' \bDelta \heta V}$, $\mbf{S}_{\bF' \bDelta \heta V}$, are not translationally invariant and can therefore be very hard to handle numerically. 

If we now assume that we average over an infinite number of disorder configurations, the reference system's propagators will be translationally invariant, since due to the translational invariance of the uncorrelated disorder probability distribution of Eq.\ (\ref{eq:uncor_P}) all values $\eta_{ij}$ will occur with the same weights for each pair of sites $(i,j)$ with the same distance $i-j$, i.e.
\begin{equation}
  \bar{\bG}_{\bF' \bDelta P V}(x_i,x_j,\tau-\tau') = \bar{\bG}_{\bF' \bDelta P V}(x_i-x_j,\tau-\tau')
  \, ,\nonumber
\end{equation}
with a translationally invariant condensate
\begin{equation}
\bar{\bP}_{\bF' \bDelta P V}(x_i)=\bar{\bP}_{\bF' \bDelta P V}(x_j) \, ,\nonumber
\end{equation}

According to Eq.\ (\ref{eq:D1}) this implies that also the average self-energies will be translationally invariant with
\begin{equation}
  \bar{\bS}_{\bF' \bDelta}(i\omega_n,k) = \bG_{\bDelta 0 0}^{-1}(i\omega_n,k) - \bar{\bG}_{\bF' \bDelta P V}^{-1}(i\omega_n,k) \, ,\nonumber
\end{equation}
and
\begin{equation}
 \bar{\mbf{S}}_{\bF' \bDelta }(x_i)  =\bF'(x_i)-\bG_{\bDelta 0 0}^{-1}(i\omega_0,k=0)\bar{\bP}_{\bF' \bDelta P V}(x_i)
 = \bar{\mbf{S}}_{\bF' \bDelta }(x_j)
 \, ,
\end{equation}
Finally,  $\Omega_{\bF' \bDelta P V} =\langle\Omega_{\bF' \bDelta \heta V}\ranp $ can be computed directly from averaging over the fixed-disorder systems.

As no fixed-disorder quantities are needed in order to evaluate the functional in Eq.\ (\ref{eq:SFT_functional}), the evaluation of the self-energy functional has now the same complexity as the disorder-free case of Ref.\ \cite{BSFT}, where the self-energies and propagators were translationally invariant by definition. The only difference lies in the treatment of the reference system, which has to be averaged over all disorder configurations $\heta$.

% --------------------------------------------------------------------
\section{Lattice observables}
\label{sec:Dis_Lattice}
% --------------------------------------------------------------------

Once a stationary solution fulfilling Eq.\ (\ref{eq:stat_FD}) has been found, the corresponding lattice observables can be computed using the self-energies
\begin{equation}
  \bS_{\heta}  \approx\bS_{\bF' \bDelta\heta V}, \quad
  \mbf{S}_{\heta}  \approx\mbf{S}_{\bF' \bDelta\heta V},
  \quad
  \bar{\bS}  \approx\bar{\bS}_{\bF' \bDelta}, \quad
  \bar{\mbf{S}}  \approx\bar{\mbf{S}}_{\bF' \bDelta}.
\end{equation}
In particular, the disorder-averaged propagators $\bar{\bG}$ and $\bar{\bP}$ can be  computed using the self-energies $\bar{\bS}$ and $\bar{\mbf{S}}$ and the free propagator $\bG_{\mbf{t}00}$ in the Dyson equations of Eq.\ (\ref{eq:D1}). The fix-disorder propagators $\bGeta$ and $\bPeta$, on the other hand, can be computed using $\bS_{\heta}$ and $\mbf{S}_{\heta}$  and the free propagator $\bG_{\mbf{t}\heta0}$ in Eq.\  (\ref{eq:D1}). As the latter are not translational invariant, however, they can only be computed on a finite sized lattice, as Eq.\  (\ref{eq:D1}) requires the inversion of a matrix in position space. It is therefore preferable to use the translationally invariant averaged propagators $\bar{\bG}$ and $\bar{\bP}$ in the thermodynamic limit.

As the arithmetic averaging is a linear operation, disorder-averaged observables of the lattice system which can be expressed as linear terms of one- and two-point quantities without any disorder-dependent prefactors, can be directly evaluated from the average propagators $ \bar{\bPhi} $ and $\bar{\bG}$. This is trivially the case for the disorder-averaged condensate through Eq.\ (\ref{eq:phi_funct_P}), while for the particle density we have 
  \begin{eqnarray}
    n &= \frac{1}{2\beta L} \left\langle
    \Tr[-\bGeta] + \bPeta^\dagger \bPeta \right\rangle_P =
    \frac{1}{2\beta L} \left(
    \Tr[-\langle \bGeta \rangle_P] + \langle \bPeta^\dagger \bPeta \rangle_P \right) \nonumber
    \\ &=
    \frac{1}{2\beta L} \left(
    \Tr[- \bar{\bG} ] + \bar{\bPhi}^\dagger \bar{\bPhi} \right)
    \, , \label{eq:latt_dens}
  \end{eqnarray}
  where we have used Eq.\ (\ref{eq:G_funct_P}) in the last step and the same definition of the trace operator $\Tr$ as in Ref.\ \cite{BSFT}.
  The same is true for the kinetic energy
  \begin{equation}
    E_{\rm{kin}} =
    \frac{1}{2\beta L}
    \left\langle
      \Tr \left[ \mbf{t} \left( \bGeta - \bPeta^\dagger \bPeta \right) \right]
    \right\rangle_P =    \frac{1}{2\beta L}
    \Tr \left[ \mbf{t} \left( \bar{\bG} - \bar{\bPhi}^\dagger \bar{\bPhi}
      \right) \right]
    \, . \label{eq:latt_ekin}
  \end{equation}

The interaction energy, on the other hand, cannot be directly evaluated from the averaged propagators as \cite{BSFT}
\begin{equation}
  E_{\rm{int}}
  =
  \frac{1}{L}
  \frac{U}{2}\sum_i
  \left\langle \langle n_i^2-n_i \rangle_{\heta} \right\rangle_P =
  -\frac{1}{4\beta L}
  \left\langle
  \Tr[\bS_{\heta}\bGeta]
  \right\ranp
  \neq
  -\frac{1}{4\beta L}\Tr[ \bar{\bS}\bar{\bG}]
  ,
\end{equation}
However, as the SFT functional is equal to the free-energy at stationarity, we have direct access to the disorder-averaged free energy of the lattice $\Omega_{\bF \mbf{t} P V}$, from which we can compute the interaction energy by the numerical derivative
\begin{equation}
E_{\rm int}=\frac{U}{L}\frac{\partial\Omega_{\bF \mbf{t} P V}}{\partial U}.\nonumber
\end{equation}

%\rmcolor{red}{Discuss implications (\ref{Self-energy_approximation1})}

% --------------------------------------------------------------------
\section{Poles in the connected Green's function}
\label{app:pole}
% --------------------------------------------------------------------

The arithmetically averaged connected Green's function of the lattice, $\bar{\bG}$, depends on momentum $k$ only through the non-interacting dispersion $\epsilon_k$ and can thus be parametrized in the single-particle energy $\epsilon$ as $\bar{\bG}(i\omega_n, \epsilon) = \bar{\bG}(i\omega_n, \epsilon = \epsilon_k)$. In terms of a local disorder-averaged self-energy $\bar{\bS}$ (such as the one used in the SFA3 reference system) it can be written as
\begin{equation}
  \bar{\bG}(i\omega_n,\epsilon)
  =
  \left[
  \sigma_z
  i\omega_n
   +
  \left(\mu-\epsilon\right)\mbf{1} -
  \bar{\bS}(i\omega_n)
  \right]^{-1}
  \, .
  \label{Ginverse_epsilon}
\end{equation}
The inversion in Eq.\ (\ref{Ginverse_epsilon}) results in simple poles of $\bar{\bG}$ whenever ${\rm det}\left[  \bar{\bG}^{-1}(i\omega_n,\epsilon)\right] = 0$, i.e., when
\begin{equation}
  \epsilon =
  \mu - \rm{Re}
  \left[\bar{\bS}_{00}(i\omega_n)\right]\pm A\left[\bar{\bS}, i\omega_n\right]
  \equiv
  \epsilon^p_{\pm}(i\omega_n)
  \, ,  \nonumber
\end{equation}
where
$A\left[\bar{\bS},i\omega_n\right]=\sqrt{\left|\bar{\bS}_{01}(i\omega_n)\right|^2-\omega_n^2-{\rm Im}\left[\bar{\bS}_{00}(i\omega_n)\right]^2}$ and $\bar{\bS}_{\nu \nu'}$ are the Nambu-components of the $2\times 2$ local self-energy.
In other words, the lattice Green's function $\bar{\bG}$ develops a pole if for some $i\omega_n$
\begin{equation}
  \min_k \epsilon_k
  \leq
  \epsilon^p_{\pm}(i\omega_n)
  \leq
  \max_k \epsilon_k
  \, ,\nonumber
\end{equation}
while the determinant of $\bar{\bG}$ can be expressed as
\begin{equation}
  \det \left[ \bar{\bG}^{-1}(\epsilon,i\omega_n) \right]
  =
  \left(
    \epsilon-\epsilon^p_{+}(i\omega_n)
  \right)
  \left(
    \epsilon-\epsilon^p_{-}(i\omega_n)
  \right)
  \, .
  \label{eq:detGbar}
\end{equation}

In the absence of $U(1)$ symmetry-breaking $\bar{\bS}_{01}(i\omega_n) = 0$, and $\bar{\bG}$ can only have a simple pole at $i\omega_0 = 0$, since $\epsilon^p_{+}(i\omega_0) = \epsilon^p_{-}(i\omega_0)$ and $A\left[\bar{\bS}, i\omega_n\neq 0\right]$ is always imaginary. %\rmcolor{red}{Implicit assumption of $\bar{\bS}_{00}(i\omega_0) = 0$?}
In the superfluid phase, where $\left|\bar{\bS}_{01}(i\omega_n)\right|>0$, the poles $\epsilon^p_{\pm}$ of $\bar{\bG}$ can be located at any Matsubara frequency.

However, the superfluid SFT groundstates we observe only develop simple poles at zero frequency (in specific parameter ranges).
This happens in the superfluid phase for strong disorder $D\gtrsim W$ and in the Bose glass phase close to the superfluid phase boundary, see the grey regions in Figs.\ \ref{fig:bg_swe} and \ref{fig:sf_swe}. 
  In the clean system, such a pole signals an instability towards $U(1)$-symmetry breaking and arises only in the metastable Mott insulator phase.
%
%In the clean system, such a pole arises only in the $U(1)$-symmetry-preserving solution (i.e.\ the Mott insulator) when this solution is metastable (i.e.\ when the groundstate is superfluid), signaling an instability towards $U(1)$-symmetry breaking.}
%
In the case studied here, which is no-longer homogeneous, as discussed in Sec.\ \ref{sec:results}, the pole is related to the appearance of isolated quasi-condensates on the lattice.

Remarkably, although the poles make non-local quantities such as, e.g., $n_k = -\sum_n {\rm Tr}  \bar{\bG}(i\omega_n,k)/2\beta$ diverge at certain values of $k$, the pole can be treated semi-analytically in the computation of local quantities, as we will show in the following.

The central quantity where the lattice Green's function enters in the SFT functional of Eq.\ (\ref{eq:SFT_functional}) is the trace-log term $\Tr\ln\left[-\bar{\bG}^{-1}\right]$, which -- as shown in Ref.\ \cite{BSFT} -- is only defined up to a regularization factor $C_{\infty}$ and can be evaluated as
\begin{equation}
  \frac{1}{2}\Tr\ln\left[-\bar{\bG}^{-1}\right] - C_{\infty}
  =
  \ln\left[
    \frac{\det\sqrt{-\bar{\bG}^{-1}}}{\det\sqrt{-\mbf{R}^{-1}}}
    \right]
  \, ,\nonumber
\end{equation}
%
%\rmcolor{red}{The factor $1/2$ is contained in the square root in the nominator. The equation above is not an equality, see Eq.\ (C1) and (C8) in our B-SFT paper.}
%
where $\mbf{R}$ is the regularization function
\begin{equation}
  \mbf{R}(i\omega_n) =
  \left\{\begin{array}{ll}
-i\mbf{\sigma_z} /\omega_n, & n \ne 0,\\
  -\beta\mbf{1}, & n = 0 .
  \end{array}\right.
  \, \nonumber
\end{equation}
By $ \Tr\ln\left[-\bar{\bG}^{-1}\right]= \Tr\ln\left[-\bar{\bG}^{-4}\right]/4$, we therefore can evaluate the trace-log term as
\begin{equation}
  \frac{1}{2}\Tr\ln\left[-\bar{\bG}^{-1}\right] - C_{\infty}
  =
  \frac{1}{4}
  \sum_n
  \int \!\! d\epsilon \, \mathcal{D}(\epsilon)
 \ln
  \left[Q(i\omega_n) \det \left[\bar{\bG}^{-1}(\epsilon,i\omega_n)\right]^{2} \right]
  \, ,
  \label{TrLogEps}
\end{equation}
where $\mathcal{D}(\epsilon)$ is the single-particle density of states, and $Q(i\omega_n)$ is the reguarlization function
\begin{equation}
  Q(i\omega_n) =
  \left\{\begin{array}{ll}
  \omega_n^4, & n \ne 0,\\
  \beta^{-4}, & n = 0. 
  \end{array}\right.\nonumber
\end{equation}

In order to evaluate the integral in Eq.\ (\ref{TrLogEps}) numerically, the dispersion is discretized on the energy grid $\epsilon=\epsilon_m$. Using the linear interpolation
\begin{equation}
\tilde{D}_m(\epsilon)=\frac{\mathcal{D}(\epsilon_{m+1})-\mathcal{D}(\epsilon_{m})}{\epsilon_{m+1}-\epsilon_m}\epsilon+\mathcal{D}(\epsilon_{m})\label{Lin_Int} \, ,
\end{equation}
of the density of states, and inserting the explicit expression for the determinant from Eq.\ (\ref{eq:detGbar}) gives
\begin{equation}
  \int \!\! d\epsilon \, \mathcal{D}(\epsilon)
  \ln
  \left[Q(i\omega_n) \det \left[\bar{\bG}^{-1}(\epsilon,i\omega_n)\right]^{2} \right]
  \approx 
  \sum_m
  \int_{\epsilon_m}^{\epsilon_{m+1}}
  d\epsilon \,
  I_m(\epsilon, i\omega_n)
  \, , \label{TrLogEps1}
\end{equation}
where the integrand is given by
\begin{equation}
  I_m(\epsilon, i\omega_n)
  \equiv
  \tilde{D}_m(\epsilon)
  \ln \left[
    Q(i\omega_n) \left(\epsilon - \epsilon^p_{+}(i\omega_n)\right)^2\left(\epsilon-\epsilon^p_{-}(i\omega_n)\right)^2
    \right]
  \, .
\end{equation}

If the interval $[\epsilon_{m},\epsilon_{m+1}]$ does not contain the poles $\epsilon^{p}_{\pm}(i\omega_n)$, the $m$th summand of Eq.\ (\ref{TrLogEps1}) can be straight-forwardly integrated analytically. Also in the presence of a pole, $\epsilon_{m}<\epsilon^{p}<\epsilon_{m+1}$, this term is integrable, and can be computed analytically by dividing up the interval into two pieces as
\begin{equation}
  \int_{\epsilon_m}^{\epsilon_{m+1}}
  d\epsilon \,
  I_m
  =
  \int_{\epsilon_m}^{\epsilon^p}
  d\epsilon \,
  I_m
  +
  \int_{\epsilon^p}^{\epsilon_{m+1}}
  d\epsilon \,
  I_m
  \, .\nonumber
\end{equation}
Also in the presence of poles (i.e.\ quasi-condensates) in the connected Green's function, the SFT functional therefore remains well-defined.

A central local observable that is directly computed from $\bar{\bG}$ is the density per site $n$, given by Eq.\ (\ref{eq:latt_dens}) and thereby by the sum
\begin{equation}
  n
  =
  - \frac{1}{\beta}\sum_{n} \int d\epsilon \,
  \mathcal{D}(\epsilon)\bar{\bG}_{00}(i\omega_n,\epsilon)\nonumber
 \approx
  -\frac{1}{\beta}\sum_{n,m}
  \int_{\epsilon_m}^{\epsilon_{m+1}} d\epsilon \,
  \tilde{D}_m(\epsilon) \bar{\bG}_{00}(i\omega_n,\epsilon)
  \, , \label{eq:nSum}
\end{equation}
where in the last step we have used Eq.\ (\ref{Lin_Int}). 
Using Eq.\ (\ref{eq:detGbar}) the Green's function component $\bar{\bG}_{00}$ can be expressed as
\begin{equation}
  \bar{\bG}_{00}(i\omega_n,\epsilon)
%  =
%  \frac{i\omega_n+\epsilon-\mu-\bar{\bS}_{00}(i\omega_n)}{
%    \det\left[\bGp^{-1}(\epsilon,i\omega_n)\right]}
%  \\
  =  \frac{i\omega_n-\epsilon+\mu-\bar{\bS}_{00}(i\omega_n)}{
    \left(\epsilon-\epsilon^p_{+}(i\omega_n)\right)
    \left(\epsilon-\epsilon^p_{-}(i\omega_n)\right)
  }
  \, .\nonumber
\end{equation}

Again, if $[\epsilon_{m},\epsilon_{m+1}]$ does not contain $\epsilon^{p}_{\pm}(i\omega_n)$, the $m$th summand of Eq.\ (\ref{eq:nSum}) can be integrated analytically. If a pole $\epsilon^p$ is present, the expression (\ref{eq:nSum}) is an integral over a simple pole, which however can be integrated analytically using the limit
\begin{equation}
  \int_{\epsilon_m}^{\epsilon_{m+1}} d\epsilon \,
  \tilde{D}_m(\epsilon) \bar{\bG}_{00}(\epsilon,i\omega_n)
  =
  \lim_{\gamma \rightarrow 0}
  \left(
  \int^{\epsilon_{m+1}}_{\epsilon^p + \gamma}  +  \int_{\epsilon_m}^{\epsilon^p - \gamma} \right)d\epsilon
  \tilde{D}_m(\epsilon) \bar{\bG}_{00}(\epsilon,i\omega_n)
    \, ,
\end{equation}
which ensures that the two divergent parts of the integrals cancel each other out, giving a finite result.

The same procedure can also be applied when computing the kinetic energy, which by Eq.\ (\ref{eq:latt_ekin}) is given by
\begin{equation}
  E_{\rm{kin}}
  =  -\frac{1}{\beta}\sum_{n} \int d\epsilon  \mathcal{D}(\epsilon) \epsilon \bar{\bG}_{00}(i\omega_n,\epsilon)
  \, .\nonumber
\end{equation}

\section*{References}

\bibliography{HHHM1.bib}
% ----------------------------------------------------------------------
\end{document}